\documentstyle[aps,prb]{revtex}

\newcommand{\zz}{{\cal Z}}
\newcommand{\bq}{\begin {eqnarray} }
\newcommand{\eq}{\end{eqnarray}}

\input epsf
\input rotate
\input psfig.sty

\begin{document}
\draft
%\twocolumn[\hsize\textwidth\columnwidth\hsize\csname @twocolumnfalse\endcsname
\title{Non-linear evolution of step meander during growth of a vicinal
surface with no desorption.
}
\author{F.\ Gillet, O.\ Pierre-Louis and C.\ Misbah}
\address{LSP, UJF-Grenoble 1,
BP87, F38402 Saint Martin d'H\`eres, France}
\date{\today}

\maketitle
\begin{abstract}

Step meandering due to a deterministic morphological
instability on vicinal surfaces 
during growth is studied.
We investigate nonlinear dynamics of a step model 
with asymmetric step kinetics, terrace and 
line diffusion, by means of a multiscale analysis.
We give the detailed derivation of the highly nonlinear evolution equation
on which a brief account has been given\cite{Pierre-Louis98a}.
Decomposing the model into driving and relaxational
contributions, we give a profound explanation to  the origin of
the unusual divergent scaling of step meander
$\zeta \sim 1/F^{1/2}$ (where $F$ is the incoming atom flux).
A careful numerical analysis indicates that
a cellular structure arises where plateaus form,
as opposed to spike-like structures  reported erroneously 
in Ref. \cite{Pierre-Louis98a}.
As a robust feature, the amplitude of these cells
scales as $t^{1/2}$, regardless of 
the strength of the Ehrlich-Schwoebel effect,
or the presence of line diffusion.
A simple  ansatz allows to
describe analytically the asymptotic regime quantitatively.
We show also how sub-dominant terms from multiscale analysis
account for the loss of up-down symmetry  of the  cellular structure.

\end{abstract}
\vspace{0.3in}
\vspace{-0.3cm}

\section{Introduction}
\label{s:intro}

\begin{figure}[htb]
\centerline{
\hbox{\psfig{figure=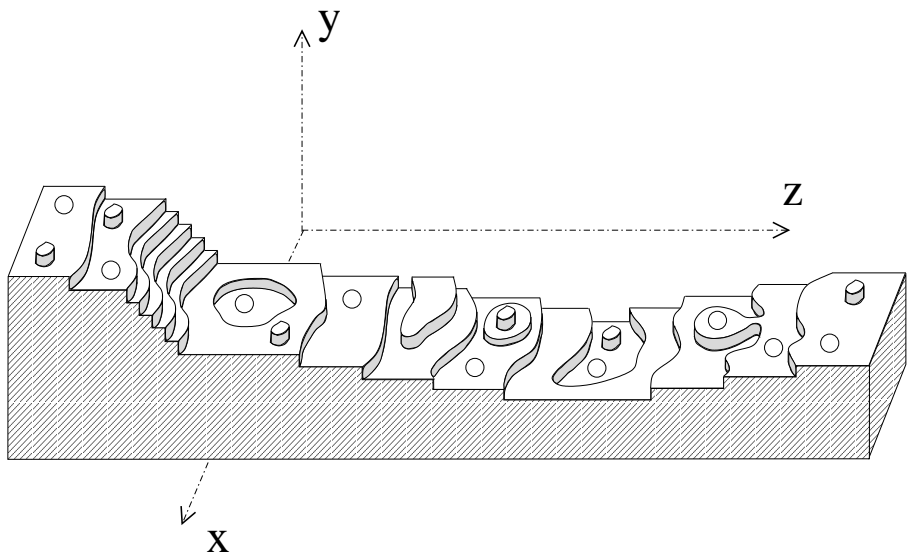,height=8cm,angle=0}}}
\caption{Schematic view of a crystal surface 
suffering various instabilities. 
Step-bunching, step-meandering, and island (or advacancy) formation
are depicted.}
\label{fig1}
\end{figure}

The production of solids by Molecular Beam Epitaxy (MBE) 
having a surface 
which is abrupt on the atomic scale
is often hampered either by a  stochastic 
roughness or due to the presence of morphological instabilities.
The stochastic roughness is often attributed to shot noise
from the incoming deposition flux.
As for determinitic instabilities, 
there are three general types of surface instabilities
leading to kinetic roughnesses: step-bunching, step meandering,
and islanding (see Fig.\ref{fig1}). The two first categories are met 
on vicinal surfaces, while the last one can either
be present on  high symmetry surfaces, a typical
mechanism  being the Ehrlich-Schwoebel effect\cite{Villain91},
or even on a vicinal surface as a secondary instability of the 
step meander\cite{Pierre-Louis98c}.

Kinetic roughening has long been  as a mystery. With regard
to MBE growth on a high symmetry surface, a prominent
example is the Kardar-Parisi-Zhang\cite{Kardar86} equation
introduced  in an attempt to describe surface
noise-induced-roughening, its one dimensional
version is: 
\bq
\partial_ty=a \Omega F+ \partial_{xx}y+\left(\partial_x y\right)^2+\eta ,
\label{KPZ}
\eq
where $y$ is the surface height, and $x$ the coordinate along the
surface(\ref{fig1}). Derivatives are subscripted, that is $\partial_ty= \partial y
/\partial t$ and so on. $F$ is the incoming flux, $a$ is the atomic height
and $\Omega$ is the atomic area. We have set the coefficients to unity,
since only the form of the equation matters in this discussion.
This  equation has given rise to a variety of investigations
both analytically and numerically (this  part has included
 analytical treatment  of the partial differential equation together with
numerical Monte-Carlo simulations which mimics KPZ dynamics). Equation (\ref{KPZ}) is phenomenological
in the sense that it is derived 
on the basis of  symmetries.

The KPZ nonlinearity is a natural candidate  in the long
wavelength limit, if  desorption
is present, or if allowance is made for  defects (such as vacancies--often
named overhangs--
in the growing solid).  
No  derivation of that equation
has been given so far, however. The reason is, in our opinion,
the lack of a continuum description of island nucleation.
In the absence of both defects and desorption (which are two usual
requirements for production of solids of interest!), the nonlinear
KPZ term is not permissible\cite{Villain91}.
In that case the equation must have a form of a conservation law, that is: 
\bq
\partial_ty=a\Omega F -{\bf \nabla}.{\bf J}, 
\label{cons_intro}
\eq
oo that upon averaging, the mean velocity is simply be given
by $a\Omega F$, as it should be; because
the KPZ nonlinearity can not be written
as a flux (i.e. as a divergence of a current) it  gives
an additional contribution
to the grwoth  velocity by a quantity
$<(\partial_x y)^2>\ne 0$ (the symbol $<..>$  stands for 
to the average), which obviously  makes no sense. There
has thus been a variety of attempts with the aim of  deriving the appropriate
surface evolution equation in that limit. Here again no 
derivation from first principles is available.

As said above, in addition to surface roughness caused by  shot noise,
nominal high symmetry as well well as vicinal
surfaces, may become inherently unstable\cite{Villain91} when
brought away from equilibrium. Nominal surfaces
may develop mounds due to the ES effect. However,
derivation of the appropriate surface evolution
equation in that case  is still a matter of debate, though
a significant progress has been achieved.

In contrast to nominal surfaces, vicinal surfaces in the step
flow regime have allowed to derive evolution equations from
first principles.
In a series of papers, we have 
shown\cite{Bena93,Pierre-Louis96,Pierre-Louis98a} that vicinal surfaces 
offer a relatively tractable situation, though often
nontrivial, where evolution equations
can be extracted from basic transport and kinetic laws.
The strategy is to first focus 
on  derivation
of step evolution equations.
Once this task is achieved, it becomes
then possible to derive the surface evolution equation.
In its general form, the evolution equation is nonlocal
and highly nonlinear.
A rather simple information is extracted, however, if we focus on 
the long-wavelength limit: that is we assume that the wavelength 
of the step meander, and thus surface modulation, is small
in comparison to the natural physical length (diffusion length
if desorption is important, otherwise the interstep distance, which
is the most frequent situation). More precisely the full growth equations,
which are highly nonlinear and nonlocal,
can be reduced to nonlinear partial differential equations, which
are more tractable and often allow a signficant analytical progress
as wil be shown here.

As shown by Bales and Zangwill\cite{Bales90}, a straight  step during MBE
growth may become morphologically
 unstable in the presence
 of  an attachment asymmetry (the Ehrlich-Schwoebel (ES) effect) at the
 step.
Close to the instability threshold, starting from the 
Burton-Cabrera-Frank (BCF)\cite{BCF51} model, we have shown\cite{Bena93} that the step
profile in the presence of desorption obeys
 the Kuramoto-Sivahinsky equation (written in a canonical form):
\bq
\partial_t \zeta=-\partial_{xx}\zeta-\partial_{xxxx}\zeta
+\left(\partial_x\zeta\right)^2 ,
\label{KS}
\eq
where $x$ is the coordinate along the step (Fig.\ref{fig1}),
and  $\zeta$ designates the step
position. In a similar fashion we have
shown later that steps on a vicinal surface  obey a set 
of coupled anisotropic Kuramoto-Sivahinsky equations\cite{Pierre-Louis96}.
The ultimate stage of surface dynamics is  found to
be spatiotemporal chaos. Two remarks are in order: (i) the KPZ nonlinearity
is of the KS type
--due to desorption-- (ii) the first term in the KS equation has
a negative sign, signalling an instability; there is a  necessity
for taking 
higher order derivatives into account in order to prevent
arbitrary short wavelength modes to develop.

A question of major importance arose 
recently\cite{Pierre-Louis98a}: if desorption is negligible,
what kind of nonlinearity should one expect? because of the conserved
character of dynamics, only terms which can be written
as derivatives of a current are allowed. We could  naively have
thought that a natural candidate would be the conserved KS equation,
namely 
$
\partial_t \zeta=-\partial_{xx}\zeta -\partial_{xxxx}\zeta
+\partial_{xx}[\left(\partial_x\zeta\right)^2]
$. 
A close inspection of the BCF equations, as shown here in details,
reveals that this is not the case, though symmetry and conservation
would dictate that form as the first plausible
candidate. We have recently shown\cite{Pierre-Louis98a} that, for
an in-phase train of steps, each step obeys  the
following nontrivial evolution equation: 

\bq
\partial_t \zeta=-\partial_x
\left[ {1 \over 1+(\partial_x\zeta)^2}
\left(\partial_x \zeta+\partial_x
\left({\partial_{xx}\zeta \over (1+(\partial_x \zeta)^2)^{3/2}}
\right)\right)\right] .
\label{PSMPK}
\eq

This highly nonlinear equation could not be inferred  from 
scaling and symmetry arguments. It is related to a singular behavior
of the amplitude of the meander that behaves as $1/F^{1/2}$
when $F$ is small.
Instead of chaos, a regular pattern is revealed, the
modulation wavelength is fixed at the very initial stages while
the amplitude of the step deformation 
follows a scaling law $w \sim t^{1/2}$.

The objective of this paper is many fold. We first
give an extensive derivation of the above evolution equation
starting from the BCF model. We shall also present a general argument
on why that singular behavior is present in the absence of desorption.
A second line of the investigation concerns higher order contributions.
It is clear that the above equation enjoys the up-down symmetry, $\zeta\rightarrow -\zeta$.
We show here that  the effect of  higher order contributions
is to destroy this up-down symmetry.
An important fact to be 
presented here is that the step profile exhibits a plateau-like morphology. This 
contradicts the preliminary simulation given in 
Ref. \cite{Pierre-Louis98a}. That simulation had suffered
from numerical inaccuracy causing spurious spikes to develop.
Finally we show that  though the full equation is highly nonlinear 
it has been possible to  provide a quantitative analytical treatment for the step morphology, and evaluate
the plateau  width  along with the meander amplitude. The results
are found to be in good agreement with numerical results.

This paper is organized as follow.
We write down the basic equations in section \ref{s:basic_eq}.
In section \ref{s:linstab} 
a linear stability analysis is performed,
which allows to evaluate  the most unstable wavelength
and the typical time for the appearance of the instability.
In section \ref{s:scal_an} we shall provide a general argument on the
extraction of the scaling of the step position with the 
incoming flux. Section \ref{s:1sided} presents the detailed derivation
of the principal evolution equation (\ref{PSMPK}) in the one-sided
limit.
We shall then
present the situation where there is a finite ES barrier. Section \ref{s:front_back} deals
with the higher order terms and their
impact on the up-down symmetry. In section \ref{s:two_sided} we generalize the derivation
of the step evolution equations to the two-sided case. 
Discussion and outlook
are  presented in section \ref{s:discuss}.

\section{Basic  equations}
\label{s:basic_eq}

We  present the model based on that of BCF, supplemneted with
asymmetric attachment kinetics as introduced by Schwoebel \cite{Schwoebel69},
and line diffusion following Ref. \cite{Ihle98}.
A vicinal surface, whose mean interstep distance is $\ell$, is considered.
On the terraces, the adatom concentration $c_m$ between steps 
$m$ and $m+1$ evolves according to:
\begin{eqnarray}
\partial_t c_m=D\nabla^2 c_m+F ,
\label{e:diff}
\end{eqnarray}
where $D$ is the adatom diffusion constant,
$F$ is an incoming flux of adatoms from a beam, and $\partial_t$
denotes the time derivative. Once an atom
is attached to the surface, it cannot detach from it (no desorption).
We consider the widely used quasistatic limit where the concentration
reaches a steady state regime on time scales much faster than  that of 
step motion. We then have to solve Eq.(\ref{e:diff}) with
the l.h.s. equal to zero.
For implications due to non-quasisteady effects
see Ref.\cite{Gillet00c}.

The excursion of  the $m^{th}$ step about 
its straight configuration  is denoted $\zeta_m(x,t)$,  so that
its position is 
$m \ell+\zeta_m(x,t)+Vt$,  where
$V$ is the mean step velocity.
We consider the case
where no step overhang is present, so that the function $\zeta(x)$ is univocal.
On both sides ($+$ and $-$ designate the lower terrace and the upper
one respectively) of step $m$,
the normal diffusion flux is linearly
related to departure from equilibrium with  kinetic coefficients
$\nu_\pm$:
\begin{eqnarray}
D \partial_n c_{m}|_+ & = & \; \; \; \nu_+ ( c_{m} -c_{eq})|_+ \nonumber\\
D \partial_n c_{m-1}|_- & = & -\nu_- ( c_{m-1} -c_{eq} )|_- ,
\label{e:cin}
\end{eqnarray}
\noindent where $c_{eq}$ is the local equilibrium concentration,
and $\partial_n$ denotes the derivative in the direction
which is normal to the step. More precisely $\partial_n\equiv {\bf n}.
{\bf \nabla}$ where ${\bf n}=(-\zeta_x,1)/\sqrt{1+\zeta_x^2}$ is 
the unit vector normal to the step,
and ${\bf \nabla}$ is the two dimensional gradient operator: ${\bf \nabla}
=(\partial_x,\partial _z)$ where $x$ is the coordinate along
an originally straight step, and $z$ the one orthogonal to it.
The attachment lengths on both sides of the steps
will be used later: $d_+=D/\nu_+$ and $d_-=D/\nu_-$. 
If $c_{eq}^{0}$ is the adatom concentration close to a straight step,
the concentration for a curved step is given by\cite{Bena93}:
\begin{eqnarray}
c_{eq} & = & c_{eq}^{0} ( 1 + \Gamma \kappa_m) ,
\label{ceq}
\end{eqnarray}
where $\Gamma=\Omega \tilde{\gamma}/k_BT$ 
(the definition of  $\Gamma$ is slightly different
from that of Ref. \cite{Pierre-Louis98b}) 
with $\tilde{\gamma}$ the step stiffness,
and $\kappa_m$, the step curvature is given by:
\begin{equation}
\kappa_m = -{\partial _{xx} \zeta_m\over [1+(\partial _x \zeta_m)^2]^{3/2}} .
\label{curv}
\end{equation}
Here  for simplicity we disregard step-step elastic interaction.
We shall come back to this point in the discussion.

At the steps, mass conservation, in the limit where the adatom
concentration is much smaller than that of the solid $1/\Omega$, imposes:
\begin{equation}
V_n =\Omega\left( D \partial_n c_m|_+ - D \partial_n c_{m-1}|_- \right)
+a \partial_s [D_L \partial_s (\Gamma\kappa_m)] ,
\label{e:cons}
\end{equation}
where $a$ is an atomic distance. Using Einstein's relation, 
the macroscopic diffusion constant
along steps is defined as $D_L=D_{st}ac_{st}$, where
$D_{st}$ and $c_{st}$ are the diffusion constant and the
concentration of mobile atoms along the step, respectively.
This expression is in agreement with that of Mullins \cite{Mullins57}.
One may object that $c_{st}$ is in fact not well defined
along a step. We shall therefore use a more general expression
derived from the Kubo formula \cite{Villain95,Giesen-Seibert93,Giesen-Seibert95}:
$D_L=a^2/\tau_L$ where $\tau_L$ is the characteristic time for detachment
of an atom from a kink.
Non-equilibrium effects related to line diffusion are not
considered in this expression.

Two sources of nonlinearities can be 
identified. The first one is  apparent in the boundary conditions (\ref{ceq},
\ref{e:cons}) because both the normal to the step and the curvature
(see Eq.(\ref{curv})) are nonlinear functions of the step profile.
The second one originates the free boundary character 
(Stephan problem) is a hidden source of nonlinearity: the concentration
field on a terrace --which is a nonlinear function of the position--
 depends on the step profile, leading  thus to a nonlinear 
concentration field as a function of the step position.

In addition to elastic interactions (not included here), 
steps are coupled via  adatom diffusion.
Dynamics are nonlocal in space and time.
With the help of an integral formulation
of the model equations, we have made explicit this nonlocality
in a previous work \cite{Pierre-Louis98c}. The use
of the quasistatic approximation suppresses delay effects,
whereas  spatial nonlocality persists.

\section{Linear stability analysis}
\label{s:linstab}

The linear stability analysis is the first step in any stability
problem. Moreover it will allow us to prepare some preliminaries
for the nonlinear analysis.
Let us define the Fourier transform of the meander as:
\begin{eqnarray}
\zeta_{\omega k\phi}= \sum_{m=-\infty}^\infty
\int_{-\infty}^{+\infty}
%\hspace{-0.5 cm}  
\int_{-\infty}^{+\infty}
%\hspace{-0.5 cm}
\,\, \zeta_m(x,t) \; e^{-i \omega t -i kx-i\phi m} \; dx \; dt,
\label{e:fourier}
\end{eqnarray}
where $i\omega$ is the pulsation of the perturbation of 
wavevector $k$ and phase shift between two neighboring
steps, $\phi$. The phase varies 
between $0$ and $2\pi$. Let us quote two special
cases. The in-phase mode $\phi=0$, corresponds
to the case where all step meanders are identical, i.e.
$\zeta_m(x,t)=\zeta_{m'}(x,t)$ for any $m$, $m'$.
The out of phase mode $\phi=\pi$ corresponds to the situation
$\zeta_m(x,t)=-\zeta_{m+1}(x,t)$. 

The derivation of the full dispersion relation can be performed
in this case along the same lines as in Ref.\cite{Pimpinelli94}. We shall
not repeat here the calculation, but give directly the result.
The quantity $i\omega$ is complex, and let us discuss separately 
the real and imaginary parts.
The real part of $i \omega$ 
takes the form
\begin{eqnarray}
\Re e(i\omega) &=&
\Omega F \frac{q}{\cal D } \left(\frac{d_--d_+}{\ell+d_-+d_+}
\right)   
\left[  (d_-+d_+) \left( q \ell \; \sinh(q \ell) - \cosh(q \ell) +
\cos (\phi) \right) + \frac{\ell}{2} q \ell \; \sinh(q \ell)  \right] \nonumber \\
& & - \Gamma q^2 \left[D_S            
\frac{q}{{\cal D}} \left( 2 \left(\cosh(q \ell)-\cos (\phi)
\right) +q(d_++d_-) \; \sinh(q \ell) \right) 
+ aD_L  q^2 \right] ,
\label{e:disp_re}
\end{eqnarray}
with $q=|k|$, and
\bq
{\cal D} & = & (d_++d_-)q \cosh(q \ell) + (d_+d_- q^2 + 1) \sinh(q \ell) .
\eq
Both macroscopic diffusion constants (adatom tracer diffusion constant $D$ times coverage of mobile atoms) on the terraces
%
%\bq
$
D_S=D \Omega c_{eq}^0
$
%\label{e:D_S}
%\eq 
%
and $D_L$ along the steps enter this relation.
The "bare" (tracer) diffusion constant of adatoms on terraces does
not appear alone. 

A positive $\Re e(i\omega)$ is a signature
of an instability.
The straight step is unstable 
during growth provided that  a normal  ES effect is present ($d_->d_+$).
Moreover, the most unstable mode is the in-phase mode
$\phi=0$. This remark will be exploited later.

The imaginary part of $i\omega$ describes propagative effects:

\begin{equation}
\Im m(i\omega) = \Omega F \sin(\phi)  \frac{q}{\cal D} (\ell+d_++d_-) .
\label{e:disp_im}
\end{equation}

The origin of this term is quite transparent. In the limit
of  a straight step ($q=0$),  we have $\Im m(i\omega) =
\Omega F \sin (\phi )$, so that the perturbed solution
takes the form  (ignoring
the real part of $i\omega$),
$\zeta_m\sim e^{in \phi+ it\Omega F \sin (\phi )} = 
e^{i\phi [n + t (V_0/\ell) \sin (\phi )/\phi ]}$. Here
we have introduced the step velocity of the uniform
train, $V_0=\Omega F \ell$. This means
that in order to travel a distance $n\ell$, it takes for a perturbation   a time
given by    $(n\ell/V_0)  (\phi/\sin (\phi ) )$. 
Since $\phi/\sin (\phi )>1$, that time is always
longer than that needed for a uniform train ($\phi=0$) to travel
the same distance. In other words, all perturbations (except the in phase
one)  travel forward
slower  that the train velocity $V_0$.
This means
that perturbations are advected backwards in the reference frame
with velocity $V_0$.

\section{Scaling analysis}
\label{s:scal_an}

Once the instability threshold is reached, any perturbation
will amplify exponentially in the course
of time, so that  nonlinear effects can no longer be disregarded.
As discussed in section \ref{s:basic_eq}, the set of growth equations
is highly nonlinear and nonlocal, so that only a "brute force" numerical
 analysis would
give a general answer. Our idea is to inspect the original
equations and try to reduce legitimately the complexity. 
The key ingredient in our analysis is the identification
of a small parameter. 

\subsection{Scaling of space and time variables}
When inspecting the dispersion
relation (\ref{e:disp_re}) for an in phase train, one realizes that the band of unstable
wavenumbers extends from $q=0$ (actually this
result is traced back to translational invariance; it corresponds
to a global motion of the train) to a critical finite value $q_c$ 
(to be defined below).
We shall assume that $q\ell$ remains small in comparison
to one,  and we come back in the discussion below to the validity
of this assumption.
In that case Eq.(\ref{e:disp_re}) takes a simpler form

\begin{eqnarray}
Re[i\omega(q\ll 1,\phi=0)]=   
 {\Omega F \ell^2\over 2}{d_--d_+ \over \ell+d_++d_-}q^2
 -  (D_S\ell+D_La)\Gamma q^4 .
\label{e:disp_in_phase}
\end{eqnarray}

We have set  here $\Phi=0$, which is the exploitation
of the fact that 
the in-phase mode is the most dangerous one. We consider later
the situation where small deviations from the in-phase mode 
are taken into account. It is seen that the range
of wavenumbers with positive $i\omega$ is given by
\bq
q_c= \left({\Omega F \ell^2 f_s\over \Gamma (D_S\ell + D_La)}
\right)^{1/2} ,
\eq
where $f_s=(d_--d_+)/(\ell+d_++d_-)$ is a parameter describing the
Ehrlich-Schwoebel effect.
The demand that the small wavenumber expansion make a sense
is satisfied by requiring $\epsilon\equiv (q_c\ell)^2 \ll 1$, 
and this is precisely 
the definition of our small parameter

\bq
\epsilon  =  {\Omega F f_s\ell ^4 \over \Gamma (D_S\ell + D_La)} .
\label{epsilon}
\eq

This guarantees the long wavelength regime. It is important
to see from the very beginning whether this limit is realistic,
or is it rather academic. Experimental data are available
on vicinal surfaces of $Cu(1,1,17)$ which have recently
revealed a meandering instability during step-flow 
growth\cite{Maroutian99}. Their data entering 
the expression of $\epsilon$ which are best known are 
$\Omega F=3 \; 10^{-3} s^{-1}$, $\ell= 21.7 \AA$. 
The step stiffness can be written as
$\tilde{\gamma} \approx (k_BT/2a)\exp(E_k/k_BT)$, $E_k$ being
the kink energy.
From a simple "bond counting" argument, one can
evaluate the adatom equilibrium concentration on a vicinal surface:
$\Omega c_{eq}^0 \sim \exp(-E_a/k_BT)$ with $E_a=3E_k$.
Using the result of Ref. \cite{Giesen-Seibert93,Giesen-Seibert95} from step fluctuations
at equilibrium, we have $E_k=0.13 eV$. 
The diffusion constant on terraces takes the form 
$D=a^2\nu_0\exp(-E_D/k_BT)$,  $\nu_0\sim 10^{13}s^{-1}$ is
an intrinsic frequency, and 
$E_D \approx 0.45 eV$ \cite{Stoltze94}.
With a lattice constant of $2.55 {\rm \AA}$, we find:
$D_S\ell= 1.4 \times 10^{15} \exp(-0.84 {\em eV}/k_BT)$.

Using the Kubo formula \cite{Villain95}, one can evaluate the
line diffusion constant:
$D_L=a^2/\tau_{L}$.
From experimental data \cite{Giesen-Seibert93,Giesen-Seibert95}:
$D_La=aD_{L0}\exp(-E_L/k_BT)$, with
$aD_{L0}=6.5 \times 10^{18} {\rm \AA}^2 s^{-1}$ and
$E_L =0.89 eV$.
With these values, we find that $D_S\ell/(D_La)\sim 10^{-2}$
in the experimental temperature range. This indicates that 
the stabilization of steps essentially occurs via line diffusion
in this situation.
In the one-sided limit ($d_+=0$ and $d_-\rightarrow \infty$),
and around $300K$ we find $\epsilon \sim 10^{-3}$, 
and $\lambda_c=2\pi/q_c \sim  10^3$ atomic distances.
This result implies that {\it a priori} the longwavelenggth limit
is appropriate.

The active modes 
in the instability are those for which
$q\ell \sim q_c\ell \sim \epsilon^{1/2}$, and therefore, lengthscales
of interest are those for which $x \sim \ell \epsilon^{-1/2}$.
The characteristic time of the instability  development is
given by the growth rate of the most unstable mode:

\begin{eqnarray}
t_m \sim {2\pi \ell^4 \over \Gamma (D_S\ell+D_La) }\; \epsilon^{-2} .
\label{tc}
\end{eqnarray}

This is obtained as $t_m\equiv 2\pi /\Re e [i\omega (q=q_m,\phi=0)]$,
$q_m$ being the wavevector of the most unstable mode,
related to $q_c$ by $q_m=q_c/\sqrt{2}$.
This relation
provides the scaling of the time variable
$t \sim \epsilon^{-2}$. 
Using the above data, we find that the 
instability typically develops after a growth of a thickness of  the order
of $100$ monolayers.

Before proceeding further, it is instructive to analyze briefly
the asynchronized train.
As pointed out in an earlier work \cite{Pierre-Louis98a},
the ES effect not only induces a morphological
instability of steps, but also leads to a ``diffusive repulsion"
between steps on a vicinal surface. This dynamical repulsion
will force steps to evolve in-phase.
The time needed for steps to organize  in-phase in the
unstable train is $t_\phi$, defined as the synchronization
time of the most unstable mode with wavevector $q=q_m=q_c/\sqrt{2}$ (i.e. the on
e
having the maximum growth rate):
\begin{eqnarray}
{1 \over t_\phi}
\sim
\partial_{\phi \phi}\Re e \left[\omega(q,\phi)|_{q=q_m,\phi=0}
\right] .
\end{eqnarray}
This time corresponds to the decay of a perturbation
having phase shifts of order one.
From linear dispersion relation (Eq.\ref{e:disp_re}):
\begin{eqnarray}
{t_\phi \over t_m}\sim \epsilon
\left({(\ell+d_++d_-)(D_S \ell + D_L a) \over
D_S \ell (d_-+d_++\ell/2) + D_L a (d_-+d_+)} \right) ,
\end{eqnarray}
where $t_m$ is the typical time for the instability
to develop (Eq.(\ref{tc})).
In the one-sided limit, $t_{\phi}/t_m\sim \epsilon$. Thus,
we expect  the synchronization time to be shorter than
the instability time, i.e. $t_{\phi}/t_m \ll 1$ since $\epsilon \ll 1$.
This means that steps will be synchronized in the early stage
of the instability. This justifies the fact to focus on small
phases, or possibly a vanishing phase, as we shall
assume later.
%We shall thus confine ourselves here
% to the study of the nonlinear
% behavior of a train of in-phase steps. The more general asynchronous
% train will constitute a topic of a forthcoming publication.

For small but finite $\phi$
we have

\begin{eqnarray}
\Re e[i\omega(q\ll 1,\phi \ll 1)]=    
& - & {\Omega F \over 2}{d_--d_+ \over \ell+d_++d_-}
\left((q\ell)^2-{d_++d_- \over \ell+d_++d_-}\phi^2 \right)
\nonumber \\
& - & \left( (D_S\ell+D_La)q^2 +{D_S  \over \ell+d_++d_-} \phi^2 \right)
\Gamma q^2 .
\end{eqnarray}

It is seen from the first term that for the phase shift 
to be relevant we must have $\phi\sim q\ell \sim \epsilon^{1/2}$.
This implies that the conjugate variable $m$ (the step position
along the vicinality) has the following scaling $m\sim  \epsilon^{-1/2}$
(meaning that one needs to travel a distance of that
order to detect phase modulations). In summary we have
the following scaling in Fourier space
\bq
q\sim \epsilon^{1/2}, \;\;\;\; \omega\sim \epsilon^2, \;\;\;\; \phi \sim  \epsilon^{1/2} ,
\eq
and their corresponding conjugate variables in real space 
\bq
x\sim \epsilon^{-1/2} , \;\;\;\; t \sim \epsilon^{-2}, \;\;\;\; m \sim  \epsilon^{-1/2} .
\label{xtmscale}
\eq

Beside  the instability character, the problem
involves progative effects which are related to the imaginary
part of $i\omega$.
Inspection 
of the imaginary part of the dispersion relation (\ref{e:disp_im})
in the long wavelength and small $\phi$ limit, shows
that $\Im m(i\omega) \sim \epsilon^{3/2}$. This defines a fast
timescale $\tau \sim \epsilon^{-3/2}$ related to propagative effects
--we mean faster than the time scale associated with the instability 
$\sim \epsilon^{-2}$. Since we shall mainly be interested by a
synchronized train (in which case the imaginary
part vanishes), we shall leave out this additional complication
for the moment, and postpone this question to a forthcoming work.

\subsection{Scaling of the meander}
\label{s:decomp}

In order to determine the nonlinear evolution equation,
following our previous work\cite{Bena93} in the presence 
of desorption, we could expand
all physical quantities (concentration, step position)  in power series 
of the small parameter, the leading contribution would be
of order $\epsilon^{0}$, followed (in principle, and in a regular
expansion) by $\epsilon^{1/2}$, since
this is the smallest power encountered above. 
This strategy worked out when desorption is present, but not in the
present context. The reason is that the first contribution
in the step profile turned out to be $\zeta\sim \epsilon^{-1/2}$. This
was viewed as an ansatz in our previous work \cite{Pierre-Louis98a}.
In this paper we provide an explanation of that fact on the basis
of general considerations. Without having resort to an explicit
derivation we shall show why this scaling 
is inherently linked with the non-desorption case.

For that purpose it is useful to identify two 'classes' of adatoms (of course just in
terms of a picture):
{\it Thermal adatoms} of concentration $c_T$ detach from a step, 
diffuse on terraces and re-attach to a step.
Mass transport associated to their motion induces
relaxation towards equilibrium.
{\it Freshly landed adatoms} of concentration $c_F$
have not yet been incorporated into a step. Their attachment
result in the non-equilibrium driving of the steps.
We can thus split the full
set of equations (\ref{e:diff}-\ref{e:cons}) into two pieces by writing 
the model equations in the following equivalent form:
\bq
D\nabla^2c_T&=&0 ,
\label{e:diff_T}
\\
D\nabla^2 c_F&=&-F .
\label{e:diff_F}
\eq
These fields obey the following boundary conditions at the steps:
\bq
D\partial_n c_T&=& \pm\nu_\pm(c_T-c_{eq}) ,
\label{e:BC_T} \\
D\partial_n c_F&=& \pm\nu_\pm c_F ,
\label{e:BC_F}
\eq
where the index $+$ and $-$ refer to both sides of the steps.
They are coupled only through mass conservation at the steps
\bq
V_n&=&v_F+v_T ,
\eq
where the driving contribution $v_F$ is proportional to the
incoming flux $F$:
\bq
v_F = D [ \partial_n c_{F_+} - \partial_n c_{F_-} ] .
\eq
Indeed from the equations obeyed by $c_F$ (Eqs.(\ref{e:diff_F}), (\ref{e:BC_F}))
by making the transformation $c_F\rightarrow c_F/F$ one sees that
$F$ scales out from the equations, implying thus that
$c_F$ must directly be proportional to $F$.

We can extract from $c_F$ the contribution of the uniform
train, which leads to a velocity given by $\Omega F\ell$,
plus another contribution due to step modulations which
must be compatible with conservation. 
$v_F$ is thus the sum of the mean step velocity and the divergence
of a flux $j$ that describes how mass is unequally distributed between
different steps, and different parts of each step:
\bq
v_F=\Omega F(\ell-\nabla j) ,
\label{drivingform}
\eq
with $j$, according to what is stated above, independent of $F$.

The relaxational contribution $v_T$ is a thermal
part and is obviously independent of $F$:
\bq
v_T&=&[D \partial_n c_T]^+_-+a\partial_s[D_L\partial_s\Gamma\kappa] .
\eq
Gradients of chemical potential $\mu$ are the driving force
of the relaxational contribution. Without loss of generality, 
and as  long as we deal with smooth and large scale perturbations,
the thermal part of the normal velocity can
be written with help of the Cahn-Hilliard \cite{Cahn58} equation:
\bq
v_T=\nabla [{\bf M}\nabla \mu] , 
\label{e:CH}
\eq
where ${\bf M}$ is the macroscopic mobility of the surface,
and $\mu=\Omega \tilde{\gamma}\kappa$ is the chemical potential.
The step index $m$ is omitted in this section to simplify notations.
Thus we shall from now on use the scalar mobility $M$ along $x$.
The chemical potential
is expressed as $\mu= \Omega\delta {\cal F}/\delta \zeta$,
where ${\cal F}$ is the step free energy. Thus,
if $f(\partial_x\zeta)$ is the free energy density, we have:
\bq
\mu=-{d \over dx} [f'(\partial_x \zeta)] .
\label{mu} 
\eq
The evolution equation of the step meander (i.e. when the step mean
velocity is subtracted) now reads:
\begin{eqnarray}
\partial_t\zeta=-\partial_x\left[
\Omega Fj+ M\partial_{xx}f'\right] .
\label{e:cons_formal}
\end{eqnarray}
Recall that $F$ is proportional to $\epsilon$ (Eq.(\ref{epsilon})), so that we can
set $F=\epsilon \bar{F}$, where $\bar{F}$ is of order one. On the other hand
$x=X\epsilon^{-1/2}$ (Eq.(\ref{xtmscale})), 
so that $\partial_{xx}=\epsilon \partial_{XX}$.
%
%This is obtained by combining (\ref{drivingform}), (\ref{e:CH}), and
%(\ref{mu}).
$j$, $M$ or $f'$ only depend on derivatives
of $\zeta$, 
due  to translational invariance. We
write their argument symbolically as 
$\{\partial_{x}\}=\{\epsilon^{1/2}\partial_{X}\}$ (which is taken
to mean any derivative and any power). 
Equation (\ref{e:cons_formal})
can be rewritten as:
\begin{eqnarray}
\partial_t\zeta=-\epsilon^{3/2}\partial_X\left[
\Omega \bar{F}j\{\epsilon^{1/2}\partial_X\}+ M\{\epsilon^{1/2}\partial_X\}
\partial_{XX}f'\{\epsilon^{1/2}\partial_X\}\right] .
\label{e:cons_formal2}
\end{eqnarray}
The central point lies in the fact that the small parameter $\epsilon$
appears as a common factor, in the first term it stems from $F$ while
in the second one it  originates from the second spatial
derivative.

In a regular expansion, close to the instability point we expect that the amplitude of modulation
is vanishingly small when $\epsilon \ll 1$. In reality, and this 
is the heart of the proof, due to the structure of 
the above equation, it will follow that no nonlinear term  can
enter the evolution equation, even if the amplitude were allowed to
be of order one.
There is even a stronger statement. Indeed
even if 
$\zeta=\epsilon^{\vartheta}H$ ($H$ is of order one),  with $\vartheta>-1/2$,
we show below that any nonlinear term has a vanishing contribution. For that 
purpose we expand any function noted $h$ (which represents  $j$....) 
 in a Taylor series
\begin{eqnarray}
h=h_0+h_1\epsilon^{1/2+\vartheta}(\partial_X H)+
h_2\epsilon^{1+2\vartheta}(\partial_X H)^2+{\rm h.o.t.} , 
\label{e:taylor}
\end{eqnarray}
where we have kept the smallest linear and nonlinear terms.
For example  a term like 
$\epsilon^{\vartheta+1}\partial_{XX}H
\ll \epsilon^{\vartheta+1/2}\partial_X H$.
%(because $\vartheta+1>\vartheta+1/2$).
Although our conclusion can be made at this stage,  let us be more explicit.
Setting  
$T=\epsilon^{2}t$, Eq.(\ref{e:cons_formal}) now reads:
\begin{eqnarray}
\epsilon^{\vartheta+2}\partial_TH
=-\epsilon^{\vartheta+2}\partial_X  \Bigl[
j_1\partial_XH
+\epsilon^{\vartheta+1/2} j_2(\partial_XH)^2
\nonumber \\
+\left(M_0+\epsilon^{\vartheta+1/2} M_1\partial_XH
\right)
\partial_{XX}
\left(f_1'\partial_XH
+\epsilon^{\vartheta+1/2} f_2'(\partial_XH)^2\right)\Bigr]
+{\rm h.o.t.} ,
\label{e:cons_scal}
\end{eqnarray}
Since $\vartheta>-1/2$, we have $\epsilon^{\vartheta+1/2} \rightarrow 0$ as 
$\epsilon \rightarrow 0$. Therefore, nonlinear terms
are irrelevant in Eq.(\ref{e:cons_scal}), and the full equation reduces
to 
a linear evolution equation:
\begin{eqnarray}
\partial_TH=-\left(
j_1\partial_{XX}H+M_0f_1'\partial_{XXXX}H \right) . 
\end{eqnarray}
For nonlinearities to be relevant in Eq.(\ref{e:cons_scal}),
we need $\epsilon^{\vartheta+1/2}\sim O(1)$, 
which is obtained when $\vartheta=-1/2$.
But then, the expansion performed in Eq.(\ref{e:taylor}) is  a priori
not legitimate.
Indeed, higher order terms become relevant:
$(\partial_x\zeta)^n\sim\epsilon^{n(\vartheta+1/2)}
\sim O(1)$ when $\vartheta=-1/2$ for any integer $n$. 
We therefore expect a highly nonlinear evolution
equation, as will be shown explicitly in the next section.

How concentration scales with $\epsilon$ can also be found 
using the decomposition of the concentration.
From equations (\ref{e:diff_F}) and (\ref{e:BC_F}), we have
$c_F  \sim F \sim \epsilon $. From Eq.(\ref{e:diff_T}), (\ref{e:BC_T}),
and (\ref{ceq}),
we find that 
\bq
c_T-c_{eq}^0 \sim c_{eq}-c_{eq}^0 \sim c_{eq}^0\Gamma \kappa \sim
\epsilon^{1/2} . 
\eq
Thus, $u=\Omega{c_F+c_T-c_{eq}^0} \sim \epsilon^{1/2}$, and
the concentration will be
written in the following form:
\bq
u(x,t)&=&\epsilon^{1/2} U(x,t) , 
\label{uexp}
\eq
with  $U(x,t)\sim O(1)$.
Similarly, the meander will be  written as:
\begin{eqnarray}
\zeta(x,t)&=&\epsilon^{-1/2} H(x,t) , 
\label{zetaexp}
\end{eqnarray}
where $H(x,t)\sim O(1)$.

It is important to show why in the presence of desorption the expansion
is regular, leading 
to the KS equation (\cite{Bena93}).
With desorption the evolution equation can no longer be written
in the form of a conservation law (\ref{cons_intro}).
Nevertheless,  the above mentioned decomposition
still holds in a slightly different form: 
instead of being proportional to $F$, the driving
part is  proportional to $F-F_{eq}$, where $F_{eq}$
is the incoming flux at equilibrium that counterbalances
ambiant desorption ($F_{eq}= c_{eq}^0/\tau$,  $\tau$ being
the characteristic residence time before desorption on a terrace).
Hence, instead of Eq.(\ref{e:cons_formal}), we have:
\begin{eqnarray}
\partial_t\zeta=(F_{eq}-F)g+N \; \partial_{xx}f' ,
\label{e:ncons_formal}
\end{eqnarray}
where $g$ and $N$ are functions of the derivatives of $\zeta$.
The second term of the r.h.s. is now 
the one expected for non-conserved relaxation to
equilibrium: it is directly proportional to the chemical potential
variations (See model A in Ref. \cite{Hohenberg77}, or \cite{Mullins57}). 
Linearizing this equation, we have
to take $g \approx \tilde{g}\partial_{xx}\zeta$ since the
first linear term proportional to $\partial_x\zeta$ is a propagative term,
not contributing to stabilization or destabilization (moreover,
this term, not invariant under the $x\rightarrow -x$ symmetry,
is not allowed).
We then find:
\begin{eqnarray}
\partial_t\zeta=[(F_{eq}-F)\tilde{g}+N_0f'_1] \; \partial_{xx}\zeta . 
\end{eqnarray}
The prefactor of $\partial_{xx}\zeta$ is the effective stiffness
of the step. An instability is signaled by a negative
sign of that prefactor.
This happens when $F>F_c=F_{eq}+N_0f'_1/\tilde{g}$.
The small parameter (that is the distance from the instability
threshold) is now $\epsilon'\sim F-F_c$.
Moreover it was found in Ref. \cite{Bena93,Pierre-Louis98b} that
$x\sim \epsilon'^{-1/2}$ and $t\sim\epsilon'^{-2}$. 
Defining as in the conserved case
$X=\epsilon'^{1/2}x$ and
$T=\epsilon'^{2}t$, and $\zeta=\epsilon'^{\vartheta'}H$,
with $X,T,H \sim O(1)$, Eq.(\ref{e:ncons_formal}) 
is now expanded for $F\approx F_c$ as:
\begin{eqnarray}
\epsilon'^{\vartheta'+2} \; \partial_TH
=-\epsilon'^{\vartheta'+2} \; \tilde{g} \; \partial_{XX}H
+ \epsilon'^{2\vartheta'+1} \; (F-F_{eq}) \; (\partial_XH)^2 \; g_2
\nonumber\\  
+\epsilon'^{2\vartheta'+2} \; \left[
N_0 \, f_2' \, \partial_{XX} \, (\partial_XH)^2
+ N_1 \, f_1' \, \partial_XH \, \partial_{XXX}H 
\right] +{\rm h.o.t.} . 
\label{e:scal_ncons}
\end{eqnarray}
As before, we use 
$\vartheta'>-1/2$ so that  expansion (\ref{e:taylor})
makes a sense.
It is seen from this equation that the leading nonlinear term 
is
$(F-F_{eq}) \, \epsilon'^{2\vartheta'+1} \, (\partial_XH)^2$.
It counterbalances the linear term $\epsilon'^
{\vartheta'+2} \, \tilde{g} \, \partial_{XX}H$ provided that  $\vartheta'=1$.
The nonlinear term here,
is that of the Kardar-Parisi-Zhang \cite{Kadar86} 
and Kuramoto-Sivashinsky \cite{Bena93,Pierre-Louis98c} 
equations (see Eq.(\ref{KPZ}) and (\ref{KS})).
It could not be present in the conserved case because
it cannot be written as the divergence of a flux. Moreover,
it is non-variational, and thus it must vanish at equilibrium,
as can bee seen from its prefactor $F-F_{eq}$.

It must be emphasized that  the  decomposition into an equilibrium
an a nonequilibrium part holds in the present
problem, but is not a general  property.
This does not have to be the case
out of equilibrium in general (an example  is that of step flow
or sublimation 
in the presence of electromigration,
such a decomposition between relaxation and driving parts
has not been made possible).

\section{One-sided synchronized steps}
\label{s:1sided}

\subsection{Multiscale analysis}
\label{s:multi}
In addition to synchronization, we first assume for simplicity
a one-sided limit (steps advance only thanks to atoms
from the terrace which is ahead), formally defined as 
$d_+=0$, and $d_- \rightarrow +\infty$.
In this limit Eq.(\ref{e:cin})
reduces to $c_+= c_{eq}$ (which is the Gibbs-Thomson
condition)  and $\partial c_-/\partial n=0$ (atoms do not descend
the steps).

As we have shown in the last section the meander $\zeta\sim \epsilon^{-1/2}$,
while the concentration field $u\sim\epsilon^{1/2}$, we  find
it convenient to set $\zeta=\epsilon^{-1/2} H$ and $u=\epsilon^{1/2} U$,
with $H$ and $U$ being quantities of order one.
Under the assumption
that these quantities are analytic
functions of  $\epsilon^{1/2}$, we seek solutions of the form
\begin{eqnarray}
U & = & U^{(0)} + \epsilon^{1/2} U^{(1/2)} + \epsilon U^{(1)} +
\epsilon^{3/2} U^{(3/2)} + \cdots \label{Uexp}, \\  
H & = & H^{(0)} + \epsilon^{1/2} H^{(1/2)} + \epsilon H^{(1)} +
\epsilon^{3/2} H^{(3/2)} + \cdots . 
\label{Hexp}
\end{eqnarray}
In order to make explicit the $\epsilon$ dependence,
and to deal with quantities of order 1, and according to (\ref{xtmscale}),
we set:
\begin{eqnarray}
x=\epsilon^{-1/2} X ,
\;\;\;\;\;\;\;
t=\epsilon^{-2}T
\end{eqnarray}
It is convenient to rescale
space by $\ell$ and time by $D/\ell^2$.
Performing the variable change: $\zz=z-\zeta_m(x,t)$,
mass conservation (\ref{e:diff}) on terraces reads:
\begin{eqnarray}
0 = {\rho^2} \; \; \partial_{\zz\zz}U +
\; \epsilon^{1/2} \left( \eta - 2 \partial_X H \; \; \partial_{X\zz}U
 -  \partial_{XX} H \; \; \partial_{\zz}U \right)
+ \; \epsilon \; \partial_{XX} U , 
\label{eq_dif_rescaled}
\end{eqnarray}
where  $\rho = [1+(\partial_XH)^2]^{1/2}$,
an $\eta=\epsilon \; D/\Omega F \ell^2$.
At the steps, the Gibbs-Thomson relation at $\zz=0$,
and a zero-flux condition at $\zz=1$, takes the form:
\begin{eqnarray}
U|_{\zz=0} & = & - {\cal K} , \label{eq_cin1_rescaled} \\
\rho^2\partial_\zz U|_{\zz=1} & = &
\epsilon^{1/2} \; \; \partial_X H \; \; \partial_X U \; \; |_{\zz=1} \; ,
\label{eq_cin2_rescaled}
\end{eqnarray}
where ${\cal K}=\Omega c_{eq}^0 \; \Gamma \; \partial_{XX}H/\rho^3$.

Mass conservation at the step (Eq.(\ref{e:cons}))
yields
\begin{eqnarray}
V + \epsilon^{3/2}\partial_T H = {\rho^2} \; \partial_\zz U |_{\zz=0}
-\epsilon^{1/2} \; \partial_X H \; \; \partial_XU |_{\zz=0} - \epsilon^{3/2} 
\; \partial_X \left({\beta \over \rho}
\partial_X {\cal K}\right) , 
\label{eq_cont_rescaled}
\end{eqnarray}
where $\beta=D_La/D_S\ell$.
The strategy is now to solve  equations
(\ref{eq_dif_rescaled}-\ref{eq_cont_rescaled}) in successively higher
orders in $\epsilon$. 

\begin{center}
{\bf order 0}
\end{center}

To this order, Eq.(\ref{eq_dif_rescaled}) reads:
\begin{equation}
\partial_{\zz\zz}U^{(0)} = 0  ,
\end{equation}
which is solved by $U^{(0)}=A^{(0)} \zz + B^{(0)}$. 
Equations (\ref{eq_cin1_rescaled}) and (\ref{eq_cin2_rescaled})
provide two conditions from which we get
$A^{(0)} =0$ and $B^{(0)} = -{\cal K}^{(0)}$. 
No contribution to step velocity is found to $0^{th}$ order. 
That is to say this order corresponds to the equilibrium case.

%??Since the small parameter is proportional
%??to the flux $F$, this is not surprising.

\begin{center}
{\bf order 1/2}
\end{center}

From (\ref{eq_dif_rescaled}) we find
that $U^{(1/2)}$ obeys an inhomogeneous equation on terraces:
\begin{equation}
\rho^{(0)2}\partial_{\zz\zz} U^{(1/2)} = -\eta ,
\end{equation}
whose general solution takes the form:
\begin{equation}
U^{(1/2)} = \frac{-{\cal Z}^2}{2\rho^{(0)2}} \; \eta + A^{(1/2)} {\cal
Z} + B^{(1/2)} .
\end{equation}
From boundary conditions at the steps (\ref{eq_cin1_rescaled}) 
and (\ref{eq_cin2_rescaled}):
\begin{eqnarray}
U^{(1/2)} & = & B^{(1/2)} = - {\cal K}^{(1/2)} \\
\rho^{(1/2)2} \; \partial_\zz U^{(1/2)} |_{\zz=1} &=&
\partial_X H^{(0)} \; \partial_X B^{(1/2)} .
\label{eq_same_1}
\end{eqnarray}
Integration constants are found to be:
\begin{eqnarray}
A^{(1/2)} & = & (\eta- \partial_X H^{(0)} \; \partial_X {\cal K}^{(0)})
\; / \; \rho^{(0)^{2}} , \\
B^{(1/2)} & = & - {\cal K}^{(1/2)} .
\end{eqnarray}
Mass conservation at the step (\ref{eq_cont_rescaled}) 
determines the mean step velocity.
Going back to physical variables, we find the expected result:
$V=\Omega F \ell$.

\begin{center}
{\bf order 1}
\end{center}

To this order, $U^{(1)}$ obeys:
\begin{eqnarray}
\partial_{\zz\zz}U^{(1)} & = & 
\frac{1}{\rho^{(0)^2}}  \left[ \partial_{XX} H^{(0)} \; \partial_\zz U^{(1/2)}
+ 2 \partial_X H^{(0)} \; \partial^2_{X\zz} U^{(1/2)}
\right. \nonumber \\ && \left.
 -  2  \partial_X H^{(0)} \; \partial_X H^{(1/2)} 
\; \partial_{\zz\zz} U^{(1/2)} -
\partial_{XX} U^{(0)} \right] \nonumber \\
& = & a + b \zz , 
\end{eqnarray}
whose general solution takes the form:
\begin{equation}
U^{(1)} = \frac{b}{6} \zz^3 + \frac{a}{2} \zz^2 +
A^{(1)} {\cal Z} + B^{(1)} .
\end{equation}
Once again, integration factors $A^{(1)}$ and $B^{(1)}$ are
found from boundary conditions
(\ref{eq_cin1_rescaled}) and (\ref{eq_cin2_rescaled}):
\begin{eqnarray}
U^{(1)}|_{\zz=0} & = & B^{(1)} = - {\cal K}^{(1)} ,  \\
\partial_\zz U^{(1)}|_{\zz=1} & = & 
{b \over 2}+a+A^{(1)} \nonumber \\
& = &
\frac{1}{\rho^{(0)^2}}  \left[ \partial_X H^{(0)}
\; \partial_X U^{(1/2)} \right. \nonumber \\
& + & \partial_X H^{(1/2)} \; \partial_X U^{(0)} \nonumber \\
& - & \left. 2 \; \partial_X H^{(0)} \; \partial_X H^{(1/2)}
\; \partial_\zz U^{(1/2)}
\right] |_{\zz=1} . 
\end{eqnarray} 
Finally, mass conservation (\ref{eq_cont_rescaled}) leads to the sought
after evolution equation
for $H^{(0)}$:
%
%\begin{eqnarray}
%\partial_T H^{(0)} & = & \frac{A^{(1)}}{\rho^{(0)^2}}
%+ 2 \; \partial_X H^{(0)} \; \partial_X H^{(1/2)} \; A^{(1/2)} \nonumber
%\\
%& & - \partial_X H^{(0)} \; \partial_X B^{(1/2)} - \partial_X H^{(1/2)} \;
%\partial_X B^{(0)} \nonumber \\
%& & -\partial_X \left( {\beta \over \rho^{(0)}}
%\partial_X {\cal K}^{(0)} \right)
%\label{eq_wished}
%\end{eqnarray}
%
\begin{eqnarray}
\partial_T H^{(0)} & = &
\partial_\zz U^{(1)}\rho^{(0)2}
-\partial_XU^{(1/2)} \partial_XH^{(0)}
-\partial_X \left( {\beta \over \rho^{(0)}}
\partial_X {\cal K}^{(0)} \right) .
\label{eq_wished}
\end{eqnarray}
Upon substitution of the expressions of $ U^{(1)}$ and $U^{(1/2)}$, one
realizes that 
terms containing $H^{({1/2})}$ 
cancel exactly in this expression, leading to a closed form for the
evolution equation for $H^{(0)}$:
\begin{eqnarray}
\partial_T H^{(0)} & & = - \partial_X 
\left[ \eta \frac{ \partial_X H^{(0)}}{2 \; \rho^{(0)^2} } + \left( 1 +
\beta \rho^{(0)} \right) \frac{\partial_X {\cal
K}^{(0)} }{\rho^{(0)^2}} \right] . 
\label{eq_meandre_H0}
\end{eqnarray}
Going back to physical variables, we obtain:
\begin{equation}
\partial_t \zeta = - \partial_x \left[ \frac{\Omega F
\ell^2}{2} \frac{\partial_x \zeta}{(1+(\partial_x\zeta)^2)} - 
\left( D_S \ell + D_L a (1+(\partial_x\zeta)^2)^{1/2} \right)
\frac{\partial_x (\Gamma \kappa)}{(1+(\partial_x\zeta)^2)} \right] . 
\label{eq_meandre_H0_var_phys}
\end{equation}
Besides the term proportional to $D_L$ (line diffusion constant),
this is the equation derived in Ref.\cite{Pierre-Louis98a} on which
we have given a brief account.

Introducing  the step macroscopic mobility ${\cal M}$, 
and  the chemical potential $\mu=k_BT \Gamma \kappa$, 
the evolution equation can be rewritten in a more 
 compact and enlightening form:
\begin{equation}
\partial_t \zeta = - \partial_x \left[ \frac{\Omega F
\ell_{\perp}^2}{2} \partial_x \zeta - {\cal M} \partial_s \mu \right] , 
\label{eq_meandre_mobil}
\end{equation}
where $s$ is the arclength along the steps, and 
$\ell_\bot=\ell/[1+(\partial_x\zeta)^2]^{1/2}$ is
the  distance between two neighboring steps measured along their normal (see Fig.\ref{fig2} for geometrical definitions.).
The effective step mobility reads:
\bq
{\cal M}  =  {D_S \ell_{\perp}+D_La \over k_BT} .
\label{eq_mobility}
\eq
The expected decomposition of step velocity 
(see section \ref{s:decomp}) is clearly seen here. 
The first term on the r.h.s. of Eq.(\ref{eq_meandre_mobil})
is the driving part. To this term  a simple
geometrical meaning can be assigned (see Appendix \ref{a:geom}).
The second term  is the relaxation part
with a mobility depending on the local step orientation. 
Note that the present mobility ${\cal M}$ and the one introduced
in section \ref{s:decomp}, noted $M$, differ by the scale factor $[1+(\partial_x\zeta)^2]^{1/2}$
which relates the arc-length $s$ to the Cartesian coordinate $x$.

\subsection{Numerical solution}
\label{s:num_simul}

%Equation (\ref{eq_meandre_H0_var_phys}) is highly nonlinear
%and cannot, as usually, be decomposed as the sum of  linear and
%nonlinear terms. 
%This specific form is at the origin of an
%unusual behavior.
In Ref \cite{Pierre-Louis98a}, 
numerical solution of Eq.(\ref{eq_meandre_H0_var_phys}) (without line diffusion)
was performed using a simple Euler scheme, and it was found that:
(i) A cellular structure takes place, the wavelength of which (the most unstable one) is fixed
at the initial stage of the instability and no
coarsening is seen.
(ii) The amplitude grows like $t^{1/2}$.
(iii) The shape of the cells is similar to the inverse
error function, that is to say it develops a spike-like morphology.
(iv) The meander is symmetric with respect to the 
transformation $\zeta \rightarrow -\zeta$. It has been realized meanwhile
that, though all these qualitative features were correct, the spikes
are the result of a  numerical deficiency in the original code.
In order to cure this problem we have to resort to a special 
numerical treatment. Different successful attempts have been made
but we shall here describe only the most robust numerical treatment.

\begin{figure}[htb]
\centerline{
\hbox{\psfig{figure=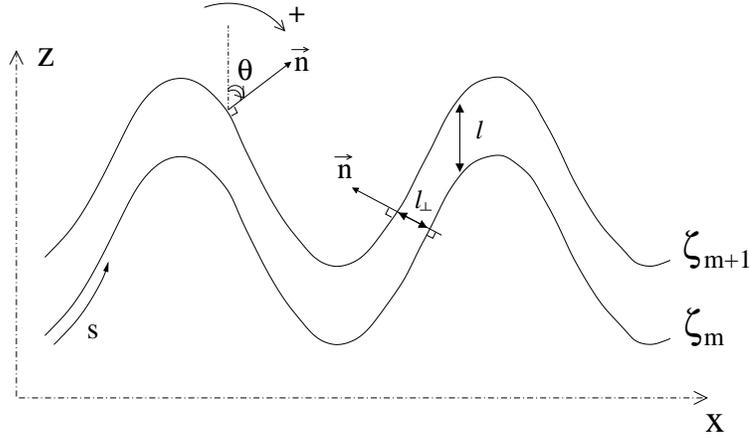,height=6.5cm,angle=0}}}
\caption{Some definitions of the notations used in the text.}
\label{fig2}
\end{figure}

%In order to avoid to deal with quantities
%having large amplitude variations,
%numerical solution of Eq.(\ref{eq_meandre_H0_var_phys}) 
%is here performed using intrinsic coordinates.
We use a powerful  geometrical representation 
of the meander \cite{Langer92,Csahok99}, in terms of 
the arclength $s$ and the angle $\theta$, oriented counterclockwise, between 
the normal and a given fixed direction (the z-axis direction) (see Fig.\ref{fig2}). 
$\theta$ is  related to $\zeta$ via: 
$\tan(\theta) = - \partial_x \zeta$ and the curvature 
simply reads: $\kappa = \partial_s \theta$.

Simple differential geometry\cite{Csahok99} 
provides us with the evolution equation for $\theta$, 
as a function of tangential and normal velocities, $v_t$ and $v_n$:
\begin{equation}
\frac{\partial \theta}{\partial t} = v_t \kappa - \frac{\partial v_n}{\partial s} .
\end{equation}
Physics is invariant under a change of
definition of the arclength $s$. This allows an arbitrary 
time-dependent re-parameterization of the curve.
This "gauge" can be seen as an additional tangential
velocity,\cite{Langer92,Csahok99} with no physical relevance. 
A particular choice that is convenient here is
the one that keeps the relative 
arclength $s/L$ constant in the course of time, 
where $L$ is the total length of the curve. 
This will ensure that the discretization points remain equally 
spaced along the curve.   
The tangential velocity reads\cite{Langer92,Csahok99}:
\begin{equation}
v_t = \frac{s}{L} \int_{0}^{L}\kappa v_n ds' - \int_{0}^{s}\kappa v_n ds' .
\end{equation}
The evolution equation of the meander (\ref{eq_meandre_H0_var_phys}) 
allows one to write the step normal velocity:
\begin{equation}
v_n = - \partial_{s} \left[ \cos(\theta) \; \sin(\theta) + 
\left( \frac{\beta +  \cos(\theta)}{\beta + 1} \right) 
\; \partial_{s} \kappa \right]  ,
\end{equation}
where time $t$ is rescaled by
$4 \ell^{4} / \epsilon^2 \Gamma (l\; D_S + a \; D_L)$
and spatial variables  $x,\zeta,$ by $\sqrt{2} \ell / \sqrt{\epsilon}$, so
that  only one parameter survives: $\beta=D_La/D_S\ell$.

Derivatives along the arclength $s$ 
are evaluated using a centered finite difference method.
We use a backward differentiation scheme
with variable step for time integration. 
This "solver" enjoys 
rather good precision and 
L-stability (that is to say it is unconditionally stable and 
optimally attenuates high-frequency (i.e. noise) 
components of the solution) that makes it 
well fitted to our specific problem. 

\begin{figure}[htb]
\centerline{
\hbox{\psfig{figure=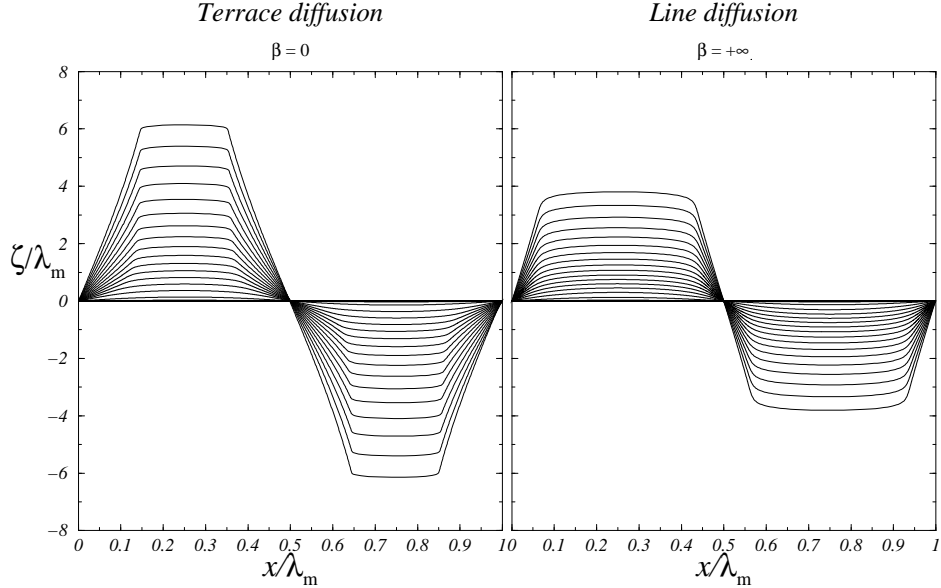,height=8cm,angle=-90}}}
\caption{Numerical solution of Eq.(64).
Meander over one period. 
Cells are symmetric and develop plateaus.
Their amplitude increases with time.}
\label{fig3}
\end{figure}

Our present simulations show qualitatively similar behavior as
that found in Ref. \cite{Pierre-Louis98a} (see Fig.\ref{fig6}). 
A major difference is revealed however: instead of spikes,
the cellular structure exhibits a plateau 
in the extrema regions \cite{rem_Krug}, as shown on Fig.\ref{fig3}.
(Here we define a plateau as a region of finite slope,
as opposed to regions where the slope diverges with time.
Hence, to some scale, plateaus are curved, and this
curvature does not tend to zero).
The width of a "plateau" reaches a constant value after a transient regime.

%This result  makes the analytical analysis made in 
%Ref. \cite{Pierre-Louis98a}, which was in favor of spikes, questionable.
%The main reason for this disagreement is the break-down
%of the validity of the variable-separation ansatz
%in the extrema region,  where $\partial_x \zeta$ is small.
%The next section reconsiders the analytical study.

\subsection{Analytical study}

\begin{figure}[htb]
\centerline{
\hbox{\psfig{figure=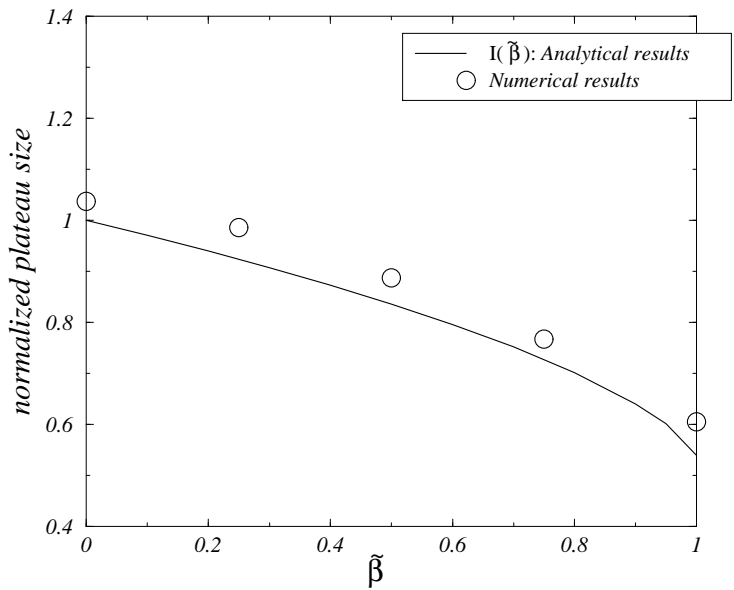,height=8cm,angle=0}}}
\caption{One sided case:
The prediction for the normalized plateau size 
$I(\tilde{\beta})$ (solid line) is compared with
numerical solution of Eq.(64) (symbols),
which corresponds to the one-sided limit.
}
\label{fig4}
\end{figure}

We show here that the above numerical results can be accounted for
using simple analytical arguments.
We give here only the main results, whereas 
details are relegated into Appendix \ref{a:late}.
The central assumption is a decomposition in two types of regions,
where two different ansatz are used. In the large slope
regions, a multiplicative variable separation  is used:
\begin{eqnarray}
\zeta_s(x,t)=A(t)g(x) ,
\end{eqnarray}
while and additive variable separation is performed for the plateau regions:
\begin{eqnarray}
\zeta_p(x,t)=B(t)+h(x) .
\end{eqnarray}
An additional constraint coming from mass conservation
allows to determine quantitatively the asymptotic behavior by matching
these two solutions. The amplitude of the meander is
found to behave as:
\begin{eqnarray}
\zeta_{max}-\zeta_{min}=2a_0t^{1/2} ,
\end{eqnarray} 
where $a_0$ is calculated
in Appendix \ref{a:late}. The rescaled meander
$\zeta(x,t)/t^{1/2}$ converges to a well defined profile,
and looks as if plateaus were formed in the extrema regions.
The width of a plateau is defined as $\lambda_0/2$.
We find:
\begin{eqnarray}
\lambda_0=\lambda_c I(\tilde{\beta}) , 
\label{e:plateau_width_1sided}
\end{eqnarray}
where $\tilde{\beta}=D_S\ell/(D_S\ell+D_La)$,
$\lambda_c=2\pi/q_c$ is the largest stable wavelength
from linear analysis, and $I$ is a function given in Appendix \ref{a:late}. Since $I(\tilde{\beta})$ decreases 
monotonously from $I(0)=1$ to $I(1)\approx 0.54$,
the  plateau size increases as line diffusion
is increased, and is always smaller than $\lambda_c/2$. 
Good quantitative agreement between the numerical solution
of Eq.(\ref{eq_meandre_H0_var_phys}) 
and these analytical predictions is found
(see Fig.\ref{fig4}).

\section{Front-Back symmetry breaking}
\label{s:front_back}

The expansion performed in section \ref{s:multi} can be
pushed to next order following the same strategy. We shall
here merely give the result and details can be 
found in Ref. \cite{Gillet00b}. Instead of a closed equation
for $H^{(0)}$, here two coupled dynamical equations for
$H^{(0)}$ and $H^{(1/2)}$ are obtained.
Going back to the physical quantity
$\zeta =\epsilon^{-1/2}( H^{(0)} + \epsilon^{1/2} H^{(1/2)})$,
the coupled equations can be recast into a single equation
for $\zeta$:
\begin{eqnarray}
\partial_t \zeta & = & - \frac{\partial}{\partial x} \left[ \frac{\Omega
\; F \; \ell_{\perp}^2}{2} \partial_x \zeta \left(1-\frac{\kappa \; \ell}{3}
\left( \frac{\ell}{\ell_{\perp}} 
+ \frac{2 \; \ell_{\perp}}{\ell} \right) \right)
- {\cal M}^{(1/2)} \frac{\partial \mu}{\partial s} \right] , 
\label{eq_meandre_H_var_phys}
\end{eqnarray}
where the macroscopic mobility of the step reads:
\begin{equation}
{\cal M}^{(1/2)} = \frac{D_S \; \ell_{\perp} + D_L \; a}{k_B T} 
 \; - \; \frac{ D_S \; \ell^2 \; \kappa}{2 \; k_B \; T} .
\end{equation}
Hence, to this order, correction to Eq.(\ref{eq_meandre_H0_var_phys})
are proportional to step curvature. 

As before, a
geometrical formulation with rescaled time and space variables, is used.
We now have two parameters $\beta$ and $\epsilon$ in the normal velocity:
\begin{eqnarray}
v_n & = & - \partial_{s} \left[ \cos(\theta) \; \sin(\theta) + 
\left( \frac{\beta +  \cos(\theta)}{\beta + 1} \right) 
\; \partial_{s} \kappa - \sqrt{\epsilon} \; \kappa \left( \frac{2}{3} \left( \cos(2\theta) 
+2 \right) \; \sin(\theta) +  \partial_s \kappa \right) \right] .
\label{eq_meandre_H_intrins}
\end{eqnarray}
The same qualitative features
as  for Eq.(\ref{eq_meandre_H0_var_phys})
are observed: the wavelength is fixed at early stages
by the one corresponding to the fastest growing mode, and the step
roughness increases with time as $t^{1/2}$
(see Fig.\ref{fig6}).
The interesting fact is the  symmetry of the shape:
the cells do not enjoy the   up-down 
symmetry, $\zeta \rightarrow -\zeta$.
Clearly, Eq.(\ref{eq_meandre_H0_var_phys}) is invariant
under the up-down symmetry $\zeta \rightarrow -\zeta$.
The symmetry breaking originates from the new terms 
as can be seen by performing the transformation $\zeta \rightarrow -\zeta$ on
Eq.(\ref{eq_meandre_H_var_phys}).
As shown in Appendix \ref{a:late}, these terms affect
the relative sizes of the  back and front plateaus,
but the $t^{1/2}$ scaling law  for the roughness  amplitude seems to
persist as 
a robust feature.
%Since higher order terms have the sign of $-\kappa$, both
%destabilizing and stabilizing terms increase in the front regions.           

The results obtained in this section are in 
complete agreement with full simulations based on a solid-on-solid
model in Ref. \cite{Pierre-Louis98a}.
Hence we have succeeded in extracting the relevant dynamics
of step meander by means of the multiscale analysis.

\begin{figure}[htb]
\centerline{
\hbox{\psfig{figure=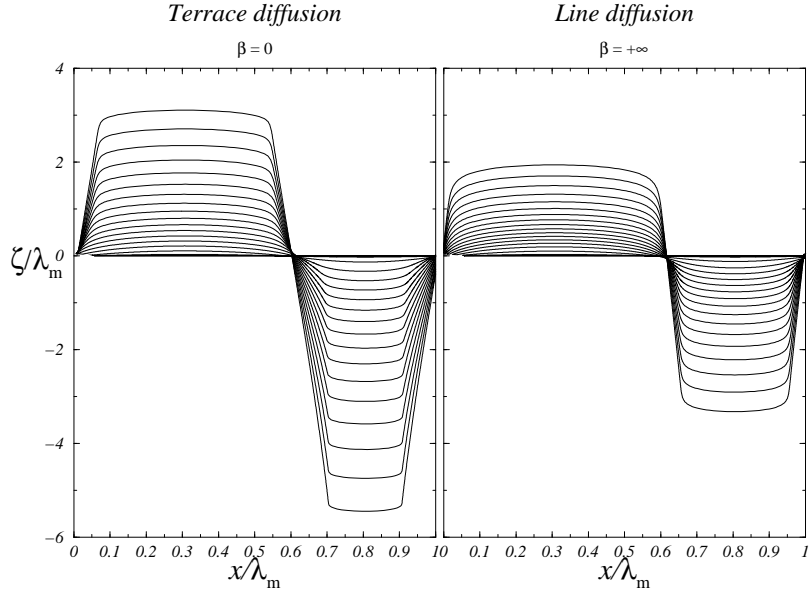,height=8cm,angle=-90}}}
\caption{Meander evolution when subdominant contributions
(to order $1/2$) are taken into account: front-back symmetry is broken.}
\label{fig5}
\end{figure}

\begin{figure}[hbt]
\centerline{
\hbox{\psfig{figure=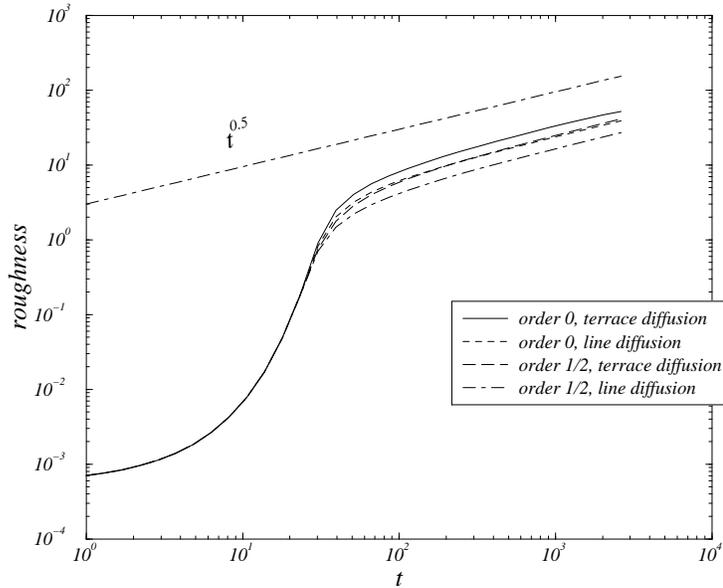,height=8cm,angle=-90}}}
\caption{One sided case:
The roughness of the meander obeys a scaling law: $w \sim t^{1/2}$.}
\label{fig6}
\end{figure}

\section{Two-sided steps in phase}
\label{s:two_sided}

In the two-sided regime (i.e. $d_+$ and $d_-$ finite),
a similar multi-scale analysis can be performed for
in-phase steps. We find:
\begin{eqnarray}
\partial_t \zeta=\partial_x\left[
-{\Omega F \over 2}
\partial_x\zeta {\ell_\bot^2(d_--d_+) \over d_++d_-+\ell_\bot}
+ {1\over (1+ (\partial_x \zeta)^2)^{1/2}}
\left(D_La+D_S  {\ell^2+\ell_\bot (d_++d_-) \over d_++d_-+\ell_\bot}\right)
\partial_x(\Gamma \kappa) \right] .
\label{e:evol_2sided}
\end{eqnarray}
Although this equation looks more complicated,
the meander evolution is qualitatively 
similar to that found in the one-sided case
(which is recovered by taking the limit $d_- \rightarrow \infty$
in Eq.(\ref{e:evol_2sided})).
Indeed, plateau formation and power law behavior
of the roughness (with the same exponent $\sim t^{1/2}$)
are also found in the two-sided case. 

\begin{figure}[hbt]
\centerline{
\hbox{\psfig{figure=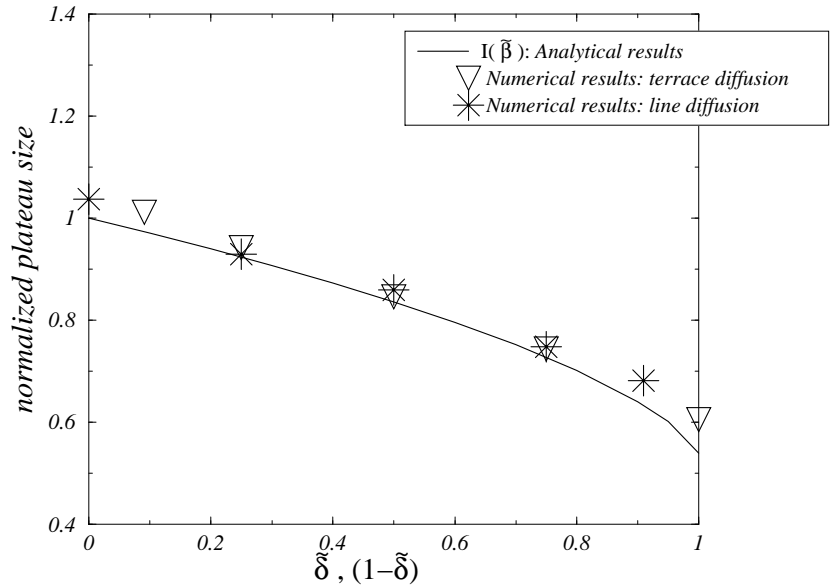,height=8cm,angle=0}}}
\caption{In solid line is plotted the normalized plateau size 
$I(\tilde{\beta})$.
Symbols represent numerical 
results: $\bigtriangledown$ refers to the pure terrace diffusion case, and 
$\star$ to pure line diffusion.}
\label{fig7}
\end{figure}

More details on step behavior, such as the plateau size, 
can be gained from the analytical investigation
of Eq.(\ref{e:evol_2sided}), as shown in Appendix \ref{a:late}.
In the pure line diffusion regime $D_La\gg D_S\ell$, we have:
\begin{eqnarray}
\lambda_0=\lambda_c I(\tilde{\delta}) , 
\end{eqnarray}
where $\lambda_0/2$ is the plateau size,
$\tilde{\delta}=\ell/(\ell+d_++d_-)$, and I is the same function
as in Eq.(\ref{e:plateau_width_1sided}). 
In the pure terrace diffusion case $D_La\ll D_S\ell$,
we find:
\begin{eqnarray}
\lambda_0=\lambda_c I(1-\tilde{\delta}) . 
\end{eqnarray}
These result are in good agreement with numerical solution
of Eq.(\ref{e:evol_2sided}) (see Fig.\ref{fig7}).

\section{Discussion and summary}
\label{s:discuss}

Starting from the BCF model, we have extracted 
a nonlinear evolution equation for the step meander.
This equation is highly nonlinear, and thus, 
could not be expected from traditional
phenomenological approaches, where linear terms
are simply supplemented with an additive nonlinear term,
as in the case of KS or KPZ equations.
%It must be emphasized however that recently
%a phenomenological study \cite{Csahok99},
%based on geometrical description of  interfaces,
%could capture  similar type of equations. Indeed, an equation which is linear
%in a geometrical representation (i.e. in terms
%of $\theta$ and $\kappa$ as in section \ref{s:num_simul}),
%can convey the impression to be of  highly nonlinear character when
%put in a Cartesian
%representation. As an example,
%the isotropic conserved equation for the normal
%velocity $v_n=\kappa+\partial_{ss}\kappa$
%found to be generic in Ref \cite{Csahok99}, reads in a Cartesian description:
%%
%\begin{eqnarray}
%\partial_t\zeta=\partial_x\left[
%{\rm arctan}(\partial_x\zeta)
%+{1 \over (1+(\partial_x\zeta)^2)^{1/2}}\partial_{x}\kappa
%\right]
%\end{eqnarray}
%%
%which shares clear similarities with Eq. (\ref{eq_meandre_H0_var_phys}).
%This shows that the geometrical
%description is in fact the natural one, and is
%more powerful than the Cartesian 
%description, owing to its  intrinsic nature.

A central   result of the present study is that
the late time power law behavior of the amplitude of the meander 
$w \sim t^{1/2}$ is a robust feature, regardless of the details
of the evolution equation (one-sided, two-sided, line diffusion....). 
It is an important task for future investigations
to see whether this property could be derived directly from the basic BCF
equations.
A second interesting point which is worth of mention is that
higher order terms destroy the up-down symmetry, but the power
law $w \sim t^{1/2}$ remains unaffected. The same conclusion
follows from full lattice gas simulations as briefly 
reported in Ref.\cite{Pierre-Louis98a}.
A natural question arises: could the amplitude temporal increase 
continue to evolve without bound in all circumstances until 
the surface breaks up into a lamellar-like pattern, or 
is there a physical mechanism, not accounted for here, 
leading to saturation of the amplitude?
Experimental observation of this instability \cite{Maroutian99}
seems to show such a saturation for the case of Cu(1,1,17), while
experiments\cite{Wu93} on Si(001) does not reveal a hint towards a saturation.
Possible candidates for amplitude saturations are (i)
strong anisotropy, (ii) elastic step interactions. We hope to
report along these lines in the near future.

%Another  issue not addressed here concerns  line diffusion.
%It has been shown recently that an asymmetry of attachment
%of step-atoms to kinks might lead to a morphological instability of the
%steps \cite{Pierre-Louis99}. This effect was not considered here.
%Incorporating nonequilibrium effects of line diffusion is important,
It is worth pointing out  that 
the use of equilibrium 
formula to evaluate the stabilizing line diffusion effect
could  be criticized, since densities of kinks and of mobile
atoms along steps depend on growth conditions \cite{Pierre-Louis99}.
%Using data from Ref. \cite{Giesen-Seibert93}, the time for 
%detachment of a kink atom is of the order of $t_1\sim 10^{-2}s$ at
%$300K$. The time between two attachment events at a kink
%resulting from the growth process is $t_2 \sim 1/(F\ell L_k)$,
%where $L_k\sim {\rm exp}[E_k/k_BT]$ is the distance between thermal kinks
%in atomic distances, with $E_k$ the kink creation energy.
%Still using numbers from Ref. \cite{Giesen-Seibert93} at 
%$300k$ and taking $\ell\sim 10$ atomic distances,
%we find that $t_2\sim 1s$ for $F\sim 10^{-3}MLs^{-1}$,
%and $t_2\sim 10^{-4}s$ when $F\sim 1MLs^{-1}$. In order to
%have local equilibration and to use thermodynamics formulas,
%we must have $t_2\ll t_1$. Clearly, it is not the case
%when growth is too fast (or temperature two low).
With regards to line mobility,
our analysis
allows  extraction of the geometry dependence of the mobility for
large meander amplitude. 
This treatment should serve as a basis for the nonlinear study 
of relaxation towards equilibrium (i.e. 
thermal smoothening) of large perturbations on vicinal surfaces \cite{Bonzel96}.

Perhaps one of the most striking result is the manifestation of 
rather 'stringent' plateaus, which are likely linked to the non-standard
character of the evolution equation.
It is interesting to note that the plateaus are a feature
of a continuum theory, a finding which is to be compared
to 
a long standing problem in the context of ES-induced mound formation
\cite{Tersoff94}.
In all previous studies, mound plateaus were indeed  considered
as a signature of the breakdown of continuum  theories.
We have shown here, in contrast, that a single equation in the
 continuum limit
can produce such plateaus,
without having resort to specific ingredients in the angular
region. It is not yet clear what kind of equation in the continuum limit would
describe these dynamics for mound formation. Is it similar or not
to the one encountered here? These questions constitute an important
line for future inquiries.
\newpage
\begin{appendix}

\section{Geometrical origin of the destabilizing term}
\label{a:geom}

In Eq.(\ref{eq_meandre_H0_var_phys}), the relaxation term 
is interpreted as a Cahn-Hilliard contribution.
We present here a derivation of the destabilizing term
from geometrical considerations in the one-sided limit.

\begin{figure}[hbt]
\centerline{
\hbox{\psfig{figure=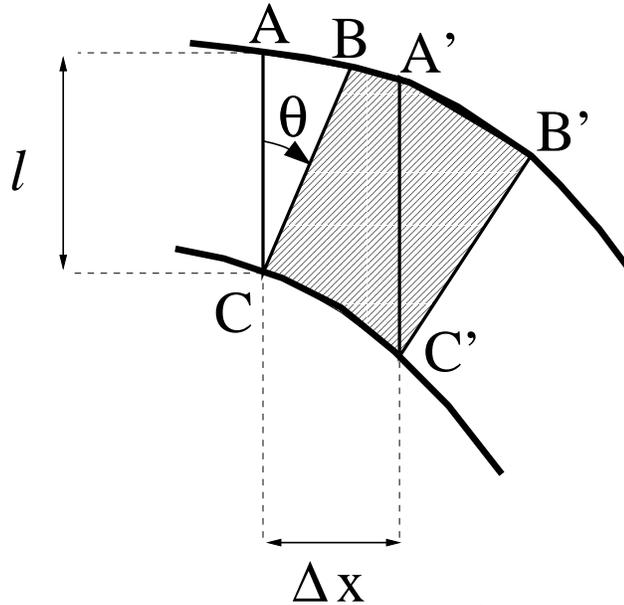,height=8cm,angle=0}}}
\caption{The element of terrace area $\Delta S$ that feeds
one element of step, is hatched. Its area (approximated by
that of BB'C'C) 
is the sum of $\ell \Delta x$, the area of AA'C'C, minus ${\cal A}(x)$
is the area of the triangle ABC, plus ${\cal A}(x+\Delta x)$
the area of A'B'C'.}
\label{fig8}
\end{figure}

Let us consider a curved part of the step as shown 
in Fig.\ref{fig8}.
In the one-sided model, step motion
results from incorporation of adatoms from the terrace ahead of it.
Mass conservation for an element of terrace surface $\Delta S$,
hatched in Fig.\ref{fig8}, reads:
\begin{eqnarray}
v\Delta x= \Omega F \Delta S + j_\bot(x)-j_\bot(x+\Delta x)
\label{e:v_dx}
\end{eqnarray}
where $v$ is the step velocity along the $z$ axis,
and  $\Delta x$ the extent of the step element $CC'$ along the $x$ axis.
The number of atoms entering the step is $v_n \Delta s=v\Delta x$. 
$j_\bot(x)$ is the total flux accross the $BC$ segment 
in Fig.\ref{fig8}. $\Delta S$ is written as:
\begin{eqnarray}
\Delta S \approx  \ell \Delta x- {\cal A}(x)+{\cal A}(x+\Delta x) ,
\label{e:area_element}
\end{eqnarray}
where $v$ is the step velocity along the $z$ axis, 
and ${\cal A}(x)$, the area of the triangle $ABC$ on 
Fig.\ref{fig8}, is a function of $\partial_x\zeta$:
\begin{eqnarray}
{\cal A}(x)={\ell^2 \over 2} \cos(\theta) \sin(\theta)=
{\ell^2 \over 2}\, 
{\partial_x\zeta \over 1+(\partial_x\zeta)^2} \; ,
\label{e:s1}
\end{eqnarray}
where $\theta$ is the angle between the $z$ axis and
the normal to the step. 
In the long wavelength limit, the local geometry of the terrace
is completely described by $\ell_\bot$, 
the length of $BC$ on Fig.\ref{fig8}, $\kappa$,
the step curvature, and their derivatives with repect to the arclength
$s$ along the steps. Since the flux $j_\bot$ can only
come from a variation of the local geometry along $s$,
we have, at  most
\begin{eqnarray}
j_\bot \sim \partial_s \ell_\bot \sim \partial_{xx}\zeta
 \ll {\cal A} \sim \partial_x \zeta \; ,
\end{eqnarray}
which shows that terms containing $j_\bot$ can be neglected
to leading order in Eq.(\ref{e:v_dx}) .
Combining Eq.(\ref{e:v_dx}), Eq.(\ref{e:area_element}), 
and (\ref{e:s1}), and letting $\Delta x$ going to zero, we find: 
\begin{eqnarray}
v=\Omega F\ell-\partial_x\left({\Omega F\ell^2 \over 2}
{\partial_x\zeta \over 1+(\partial_x\zeta)^2}\right) .
\label{e:v_geom}
\end{eqnarray}
Once the mean step velocity $V=\Omega F\ell$ is subtracted,
we recover the first term of Eq.(\ref{eq_meandre_H0_var_phys}).

\section{Late time behavior}
\label{a:late}

In this appendix we derive analytically
the main results obtained numerically.
Despite the highly nonlinear character of the evolution equation,
we show here that some simple ansatz allows to
describe the asymptotic regime with good accuracy.

\subsection{Large slope regions and extrema regions}
We found the conserved evolution equation of the step meander:
\begin{eqnarray}
\partial_t \zeta=-\partial_xj[\zeta] ,
\label{e:fluxapp}
\end{eqnarray}
with the mass flux (see Eq.(\ref{e:evol_2sided})):
\begin{eqnarray}
j[\zeta] &=& {1 \over (1+ (\partial_x \zeta)^2)^{1/2}} 
%\Bigl[
\left[
{\alpha\, \partial_x\zeta \over
\delta +(1+ (\partial_x \zeta)^2)^{1/2}}
\right.
\nonumber \\
&+& 
\left.
\left(D_La+D_S \ell
{ 1+ \delta (1+ (\partial_x \zeta)^2)^{1/2} \over 
\delta +(1+ (\partial_x \zeta)^2)^{1/2}}\right)
\partial_{xx}\left({\Gamma \partial_x\zeta \over 
(1+ (\partial_x \zeta)^2)^{1/2}}\right) 
%\Bigr]
\right] ,
\label{e:evol_2sided_a}
\end{eqnarray}
where $\delta=l/(d_++d_-)$,
and $\alpha=\Omega F \ell^2 (d_--d_+)/2(d_++d_-)$.
For the  large slope region we make use of the
 variable separation:
\begin{eqnarray}
\zeta_s(x,t)=A(t) \, g(x) ,
\label{e:var_separ}
\end{eqnarray}
where $A \gg 1$, and $\partial_xg \neq 0$ for any value of $x$. 
Substituting  in Eq.(\ref{e:evol_2sided_a}),
one finds that the destabilizing term (proportional to $\alpha$) 
dominates and:
\begin{eqnarray}
AA'=\alpha {g'' \over gg'^2}=C ,
\end{eqnarray}
where $C$ is a constant, and the prime stands for the derivative.
The late time solution of these equations reads:
\begin{eqnarray}
A &=& (2Ct)^{1/2} \\
g &=& (2 \alpha / C)^{1/2}{\rm erf^{-1}}(4x/\lambda_s) \; .
\end{eqnarray}
$\lambda_s$ being a constant, and erf$(x)$ the error function.
Inserting these expressions in Eq.(\ref{e:var_separ}), we find
that the meander does not depend on $C$:
\begin{eqnarray}
\zeta_s(x,t)=2 (\alpha t)^{1/2} {\rm erf^{-1}}(4x/\lambda_s) \; ,
\label{e:steep_zeta}
\end{eqnarray}

\begin{figure}[hbt]
\centerline{
\hbox{\psfig{figure=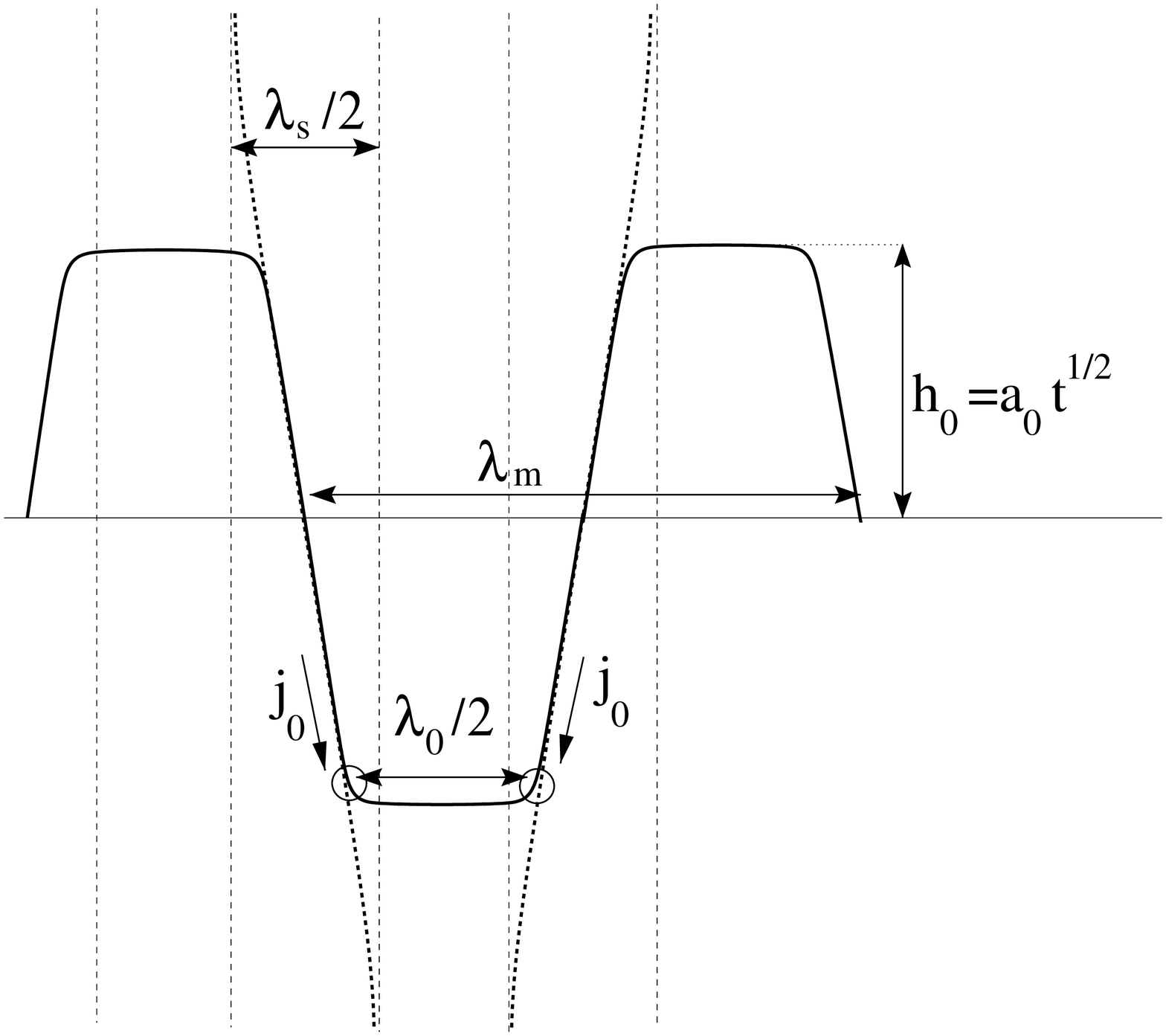,height=8cm,angle=0}}}
\caption{Asymptotic meander morphology. $\lambda_m$ is 
the period of the meander and is also 
the most unstable wavelength obtained from linear stability 
analysis. $h_0$ is the amplitude of the meander 
and $\lambda_0/2$ is the width of a plateau.
In the large slope regions (of width $\lambda_s/2$), 
the meander can be fitted by an erf like 
function (see Eq.(\ref{e:steep_zeta})). 
$j_0$ is the mass flux coming from these large slope 
regions toward the plateaus.}
\label{fig9}
\end{figure}

This solution describes regions of
large slopes, but is not expected to accurately describe the shape
around the extrema of 
$\zeta$  where the slope $\partial_x \zeta$
approaches zero.
In those  regions of width $\lambda_0/2$
and of meander amplitude of the order of $h_0$ (see Fig.\ref{fig9}),
global mass conservation implies:
\begin{eqnarray}
2 j_0= {\lambda_0 \over 2} \partial_t h_0 ,
\label{e:mass_cons_ext}
\end{eqnarray}
where $j_0$ is the mass flux coming from the large slope
regions. From Eq.(\ref{e:evol_2sided_a}), we have: 
\begin{eqnarray}
j_0 \approx {\alpha \over \partial_x \zeta_s(x_0,t)} ,
\end{eqnarray} 
where $x_0=(\lambda_m-\lambda_0)/4$ is the abscissa of the
crossover point between high slope and extrema regions, and
the period of the meander $\lambda_m$ is that of
the most unstable mode obtained from linear analysis. 
Using $h_0=\zeta_s(x_0,t)$ and Eq.(\ref{e:steep_zeta}):
\begin{eqnarray}
{\lambda_0\over 4}
={\lambda_m \over 4}-{\lambda_s \over 4}
{\rm erf}\left[{ h_0 \over 2(\alpha t)^{1/2}}\right] ,
\label{e:x0}
\end{eqnarray}
so that Eq.(\ref{e:mass_cons_ext}) now reads:
\begin{eqnarray}
{1 \over t^{1/2}}\,
{\alpha^{1/2}\lambda_s \over 4\sqrt{\pi}}\,
{\rm exp}\left[-
\left({ h_0 \over 2(\alpha t)^{1/2}}\right)^2\right]
\nonumber \\
=\partial_t h_0
\left[{\lambda_m\over 4}-{\lambda_s \over 4} 
{\rm erf}\left[{ h_0 \over 2(\alpha t)^{1/2}}\right] \right] ,
\label{e:h}
\end{eqnarray}
This equation has the trivial solution:
$h_0=a_0t^{1/2}$. Using this solution and Eq.(\ref{e:x0}),
$\lambda_0$ is seen not to depend on time.
In the extrema region, we therefore look for solutions of the form:
\begin{eqnarray}
\zeta_p(x,t)=B_\pm(t)+h(x)  .
\label{e:plane_zeta}
\end{eqnarray}
with $B_\pm(t)=\pm a_0t^{1/2}$,
where the plus and minus signs refer
to the   maxima and the minima regions respectively.
Upon substitution in the evolution equation (\ref{e:evol_2sided_a}),
we find that the problem amounts to finding the stationary
solutions $\partial_xj[h(x)]=0$. Looking for
solutions with left-right symmetry
$x \rightarrow -x$ we finally have to solve 
\begin{eqnarray}
j[h(x)]=0  .
\label{e:zero_flux}
\end{eqnarray}
This will be exploited  in the next section.

The parameters $\lambda_s$, $\lambda_0$ and $a_0$
are not independent. Using Eq.(\ref{e:mass_cons_ext}) and
Eq.(\ref{e:x0}), we have two relations,
so that  $\lambda_s$ and  $a_0$ can be determined as a function
of $\lambda_0$. From Eq.(\ref{e:h}), we get an implicit 
equation for $a_0$:
\begin{eqnarray}
{\lambda_m \over \lambda_0}-1=\sqrt{\pi}
\left({a_0 \over 2 \alpha^{1/2}} \right)
{\rm exp}\left[\left({a_0 \over 2 \alpha^{1/2}} \right)^2\right]
{\rm erf}\left[{a_0 \over 2 \alpha^{1/2}}\right] .
\label{e:a0}
\end{eqnarray}
The expression for  $\lambda_s$ is:
\begin{eqnarray}
\lambda_s=\lambda_0 \sqrt{\pi}
\left({a_0 \over 2 \alpha^{1/2}} \right)
{\rm exp}\left[\left({a_0 \over 2 \alpha^{1/2}} \right)^2\right] .
\label{e:lambdas}
\end{eqnarray}
Hence, the asymptotic behavior of the meander (defined by
Eq.(\ref{e:steep_zeta}) for high slopes and Eq.(\ref{e:plane_zeta})
for small ones)
only depends on
 the size of the extrema region $\lambda_0/2$ (since the two
 parameters $a_0$ and $\lambda_s$ are linked to $\lambda_0$).
 How the plateau size is related to the model parameters
 will be considered in the next section.

\subsection{Plateau size}
In the one-sided limit we have $\delta=0$, 
and  Eq.(\ref{e:zero_flux}) yields (in view of Eq.(\ref{e:evol_2sided_a})):
\begin{eqnarray}
0=\alpha {h' \over 1+h'^2}
+\left( {D_S\ell\Gamma \over 1+h'^2}+{D_La\Gamma \over (1+h'^2)^{1/2}}
\right)\partial_{xx} \left({h' \over (1+h'^2)^{1/2}}
\right) .
\label{e:zero_flux_onesided}
\end{eqnarray}
Introducing the abbreviation:
\begin{equation}
m= {h'\over (1+h'^2)^{1/2}} ,
\end{equation}
we can rewrite it in a familiar form:
\begin{eqnarray}
\left({(D_S\ell+D_La)\Gamma \over \alpha}\right) m"=
-{m \over \tilde{\beta}(1-m^2)^{1/2}+(1-\tilde{\beta})}=
-{dU \over dm} ,
\label{e:dyn_m}
\end{eqnarray}
analogous to that describing the motion of a particle
of position $m$ as a function of time $x$ in a potential $U$.
We have defined  $\tilde{\beta}=D_S\ell/(D_S\ell+D_La)$, and:
\begin{eqnarray}
U(m)=\int_0^m{m\, dm \over  \tilde{\beta}(1-m^2)^{1/2}+(1-\tilde{\beta})} .
\label{e:u}
\end{eqnarray}
Multiplying Eq.(\ref{e:dyn_m}) by $m'$ and integrating with respect to
$x$, we get the analogue of the "energy conservation" condition:
\begin{eqnarray}
{1 \over 2} \left({(D_S\ell+D_La)\Gamma \over \alpha}\right) m'^2+U(m)=U(m_0) ,
\label{e:energ_cons}
\end{eqnarray}
where $m_0$ is the turning point (i.e. $m'=0$ when
$m=m_0$). We look for solutions
having two "vertical" tangents where 
the step slope diverges: $\partial_x\zeta \rightarrow \pm \infty$,
(i.e. $m_0 \rightarrow \pm 1$)
in order to match the extrema solution with the 
large slope region.
The size of the extrema region reads:
\begin{eqnarray}
{\lambda_0 \over 2}=\int_0^{\lambda_0 / 2}dx=
\int^{m_0}_{-m_0}{dm \over m'} ,
\end{eqnarray}
where $m_0 \rightarrow 1$. Using 
Eq.(\ref{e:energ_cons}), we find:
\begin{eqnarray}
{\lambda_0 \over \lambda_m}= {I(\tilde{\beta}) \over\sqrt{2}} ,
\label{e:plateau_size_onesided}
\end{eqnarray}
where
\begin{eqnarray} 
\lambda_m=2\pi
\left(2{(D_S\ell+D_La)\Gamma \over \alpha}\right)^{1/2}
\label{e:wavel_onesided}
\end{eqnarray}
is the the wavelength of the most unstable
mode obtained from linear analysis, and 
\begin{eqnarray}
I(\tilde{\beta})={1 \over \pi\sqrt{2}}
\int_{-1}^{1} 
{dm \over [U(1)-U(m)]^{1/2}} 
\label{e:int}
\end{eqnarray}
is plotted in Fig.\ref{fig4}. 
$I(\tilde{\beta})$ is a decreasing function with
$I(0)=1$ and $I(1)\approx 0.54$. Hence the extrema region size is
finite and always smaller than 
$\lambda_c=\lambda_m/\sqrt{2}$.

The meander variation in this region, as compared to the total
amplitude of the meander, decreases as $t^{-1/2}$,
and the step looks as if plateaus were present.

The reader is invited to repeat the calculation in the two-sided case
--where $\delta$ is finite,
in presence of pure line  ($\tilde{\beta}=0$) or
terrace diffusion ($\tilde{\beta}=1$). Surprisingly, the same 
integral $I$ appears. Let us define 
$\tilde{\delta}=\ell/(\ell+d_++d_-)$.
We find in the pure line diffusion case:
\begin{eqnarray}
{\lambda_0 \over \lambda_m}= {I(\tilde{\delta}) \over\sqrt{2}} ,
\label{e:plateau_size_2sided_line}
\end{eqnarray}
and in the pure terrace diffusion case:
\begin{eqnarray}
{\lambda_0 \over \lambda_m}= {I(1-\tilde{\delta}) \over\sqrt{2}} .
\label{e:plateau_size_2sided_terr}
\end{eqnarray}
Hence, $\lambda$ never exceeds $\lambda_m/\sqrt{2}=\lambda_c$
the largest wavelength for which the meander is linearly stable.

We can use a similar treatment to analyze the case
with higher order terms (eq.\ref{eq_meandre_H_var_phys}).
The main point is that terms proportional to $\epsilon^{1/2}$
do not affect the long time behaviour obtained from the
ansatz \ref{e:var_separ}). Consequently, in large slope regions, we expect once again $\zeta \sim t^{1/2}$.
Terms breaking
the front-back symmetry will affect differently
maxima and minima regions, as seen in Fig.\ref{fig5},
because the effective potential $U(m)$ is not invariant under the 
$m \rightarrow -m$ transformation anymore, which corresponds
to the up-down $z \rightarrow -z$ for step meander.
We shall not develop here further this point; 
for more details see\cite{Gillet00b}.

The numerical solution of Eq.(\ref{e:fluxapp}) is performed in order
to check the validity of the analytical results.
First, the qualitative profile of the meander is 
in good agreement  with the above description.
We found the predicted scaling of the amplitude
of the meander $\sim t^{1/2}$ in all simulations performed so far,
expect in the case of very small kinetic lengths $(d_++d_-)/l < 10^{-2}$,
where we were not able to explore the late time behavior,
due to bad numerical convergence.

As shown in Fig. 4 and 7, the observed plateau size, is in very good agreement
with the prediction of Eq.(\ref{e:plateau_size_onesided},
\ref{e:plateau_size_2sided_terr},\ref{e:plateau_size_2sided_line}).

\begin{figure}[hbt]
\centerline{
\hbox{\psfig{figure=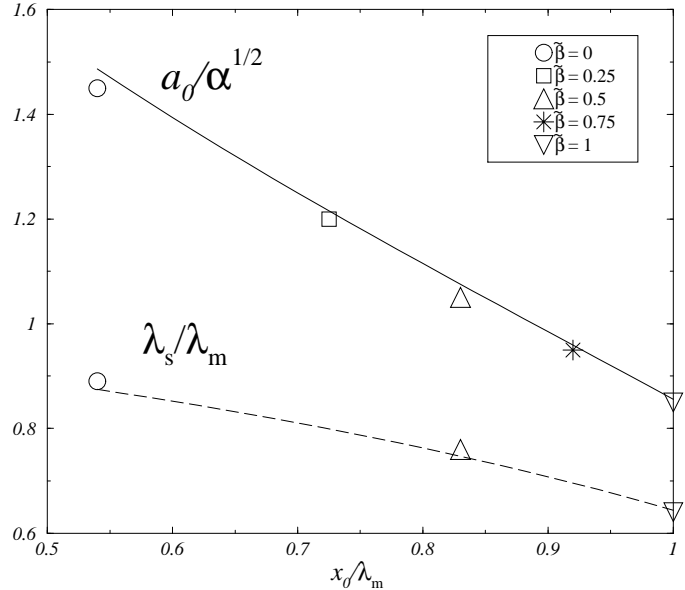,height=8cm,angle=-90}}}
\caption{The solid line represents the prediction for the amplitude $a_0$ as a 
function of $x_0/\lambda_m$. The dashed curve is the prediction
of the dimensionless ratio $\lambda_s/\lambda_m$. 
Symbols represent data from numerical solution of Eq. (64)
(the one-sided case), as $\tilde{\beta}$ is varied. }
\label{fig10}
\end{figure}

The value of $a_0$ is extracted from the evolution 
of the meander amplitude via the relation:
\begin{eqnarray}
\zeta_{p,max}-\zeta_{p,min} \approx 2 a_0 t^{1/2}  .
\end{eqnarray}
$\lambda_s$ is calculated from a fit of $\partial_x\zeta$ at
$\zeta=0$: 
\begin{eqnarray}
\partial_x\zeta_s|_{\zeta=0}=
{4\sqrt{\pi} \over \lambda_s} (\alpha t)^{1/2} .
\end{eqnarray}
In Fig.\ref{fig10}, both numerical values are compared to the prediction
of Eq.(\ref{e:a0}) in the one-sided limit, where $\lambda_0$ is calculated
from Eq.(\ref{e:plateau_size_onesided}).
Once again, good agreement is found.

\end{appendix}

%\bibliography{/home/phase4/misbah/bibgroup/sample}
%\bibliographystyle{prsty}

%------------------------------------

\end{document}